\documentclass{aa}

\newif\iffigure
\figurefalse
\figuretrue

\DeclareRobustCommand{\erase}{\bgroup\markoverwith{\textcolor{red}{\rule[.5ex]{2pt}{2pt}}}\ULon}

\usepackage[switch]{lineno}

\usepackage{graphicx,natbib,url,twoopt}
\usepackage[varg]{txfonts}           
\usepackage{hyperref}              
\usepackage{pdfcomment}              
\usepackage{acronym}                 
\usepackage{comment}
\usepackage{ulem}                    
\usepackage{cases}
\usepackage{empheq}                  
\usepackage{here}
\usepackage{bm}
\usepackage{tabularx}
\usepackage[version=3]{mhchem}
\hypersetup{
 colorlinks=true,%
 linkcolor=blue,
 citecolor=blue,
}

\usepackage[
  draft,
  authormarkup=none,
  commandnameprefix=ifneeded
]{changes}
\definechangesauthor[color=black]{R1}
\definechangesauthor[color=black]{R2}
\definechangesauthor[color=black]{R3} 
\definechangesauthor[color=black]{R4} 

\usepackage{framed}
\usepackage{mdframed}
\newcommand*\patchAmsMathEnvironmentForLineno[1]{
  \expandafter\let\csname old#1\expandafter\endcsname\csname #1\endcsname
  \expandafter\let\csname oldend#1\expandafter\endcsname\csname end#1\endcsname
  \renewenvironment{#1}
     {\linenomath\csname old#1\endcsname}
     {\csname oldend#1\endcsname\endlinenomath}}
\newcommand*\patchBothAmsMathEnvironmentsForLineno[1]{
  \patchAmsMathEnvironmentForLineno{#1}
  \patchAmsMathEnvironmentForLineno{#1*}}
\AtBeginDocument{
\patchBothAmsMathEnvironmentsForLineno{equation}
\patchBothAmsMathEnvironmentsForLineno{align}
\patchBothAmsMathEnvironmentsForLineno{flalign}
\patchBothAmsMathEnvironmentsForLineno{alignat}
\patchBothAmsMathEnvironmentsForLineno{gather}
\patchBothAmsMathEnvironmentsForLineno{multline}
}

\bibpunct{(}{)}{;}{a}{}{,}    

\makeatletter
\newcommand{\bibnote}[2]{\global\@namedef{#1note}{#2}}
\newcommand{\biblink}[2]{\global\@namedef{#1link}{#2}}

\makeatother

\makeatletter
 \newcommandtwoopt{\citeads}[3][][]{%
   \nonstopmode
   \href{http://adsabs.harvard.edu/abs/#3}%
        {\def\hyper@linkstart##1##2{}%
         \let\hyper@linkend\@empty\citealp[#1][#2]{#3}}
   \biblink{#3}{\href{http://adsabs.harvard.edu/abs/#3}{ADS}}%
   \errorstopmode}            
 \newcommandtwoopt{\citepads}[3][][]{%
   \nonstopmode
   \href{http://adsabs.harvard.edu/abs/#3}%
        {\def\hyper@linkstart##1##2{}%
         \let\hyper@linkend\@empty\citep[#1][#2]{#3}}
   \biblink{#3}{\href{http://adsabs.harvard.edu/abs/#3}{ADS}}%
   \errorstopmode}            
 \newcommandtwoopt{\citetads}[3][][]{%
   \nonstopmode
   \href{http://adsabs.harvard.edu/abs/#3}%
        {\def\hyper@linkstart##1##2{}%
         \let\hyper@linkend\@empty\citet[#1][#2]{#3}}
   \biblink{#3}{\href{http://adsabs.harvard.edu/abs/#3}{ADS}}%
   \errorstopmode}            
 \newcommandtwoopt{\citeyearads}[3][][]{%
   \nonstopmode
   \href{http://adsabs.harvard.edu/abs/#3}%
        {\def\hyper@linkstart##1##2{}%
         \let\hyper@linkend\@empty\citeyear[#1][#2]{#3}}
   \biblink{#3}{\href{http://adsabs.harvard.edu/abs/#3}{ADS}}%
   \errorstopmode}            
\makeatother

\newacro{ADS}{Astrophysics Data System}
\newacro{NLTE}{non-local thermodynamic equilibrium}
\newacro{NASA}{National Aeronautics and Space Administration}

\begin{document}
\authorrunning{A. Kuwahara et al.}
\titlerunning{Cooling-regulated gas accretion onto gap-opening planets}
   \title{Cooling-regulated gas accretion onto gap-opening planets}
      \author{Ayumu Kuwahara\inst{1} 
          \thanks{\email{ayumu.kuwahara@sund.ku.dk}} 
          \and Michiel Lambrechts\inst{1}
          \and Minghao Zhang\inst{2,3}
          \and Ruobing Dong\inst{4}
          \and Hideko Nomura\inst{5,6}}

   \institute{Center for Star and Planet Formation, \added[id=R1]{Globe} Institute, University of Copenhagen, Øster Voldgade 5-7, 1350 Copenhagen, Denmark
        \and
             National Astronomical Observatories, Chinese Academy of Sciences, Beijing 100012, People’s Republic of China
        \and
             School of Astronomy and Space Science, University of Chinese Academy of Sciences, Beijing 100049, People’s Republic of China
        \and
             Kavli Institute for Astronomy and Astrophysics, Peking University, 5 Yiheyuan Road, Haidian District, Beijing, 100871, People’s Republic of China
        \and
             National Astronomical Observatory of Japan, 2-21-1 Osawa, Mitaka, Tokyo 181-8588, Japan
        \and
             Department of Astronomical Science, The Graduate University for Advanced Studies, SOKENDAI, 2-21-1 Osawa, Mitaka, Tokyo 181-8588, Japan
         }

   \date{Received September XXX; accepted YYY}

 
  \abstract{
    Gas accretion onto forming planets controls the final masses of giant planets and provides observable signatures of ongoing formation.
    How this process depends on \added[id=R1]{the cooling properties of these newly attracted gas remains} poorly constrained.
    We present long-term, three-dimensional global hydrodynamical simulations to quantify gas accretion onto gap-opening planets \added[id=R1]{in the mass range between 1 and 3 Jupiter masses.} 
    We systematically vary the cooling time, \added[id=R1]{$\beta$, from near-isothermal ($\beta=10^{-2}$ in units of orbital time) to near-adiabatic ($\beta=10^{2}$)}, and follow the evolution until a quasi-steady state is reached. 
    \added[id=R1]{Our simulations show} that the gas accretion rate decreases monotonically with increasing $\beta$, as \added[id=R4]{$\dot{M}_{\rm acc}\propto\beta^{-0.18}$,} reaching values at $\beta=10^2$ that are approximately an order of magnitude lower than locally isothermal predictions, largely independent of planet mass.
    The reduction in accretion is traced to thermodynamic restructuring of the circumplanetary region: inefficient cooling weakens shocks, narrows the accretion bands feeding the circumplanetary disk. 
    \added[id=R2]{Our results imply that thermodynamic effects should be taken into account when interpreting observed accretion rates of young planets, and may introduce systematic uncertainties in commonly used locally isothermal assumptions.}
}
   

   \maketitle


\section{Introduction}\label{sec:Introduction}
In the core-accretion paradigm, planets form in protoplanetary disks: embryos first grow by accreting solids and, once sufficiently massive, bind nebular gas to form an envelope in hydrostatic equilibrium \citep[e.g.,][]{pollack1996FormationGiantPlanets}.
As the envelope cools and contracts, it can no longer maintain hydrostatic balance once the core exceeds a critical mass of approximately $10\,M_\oplus$ (Earth masses), triggering runaway gas accretion and rapid growth of the planet into a gas giant \citep{mizuno1980FormationGiantPlanets,ikoma2000FormationGiantPlanets}.
The gas accretion rate is a key quantity that regulates the final masses of giant planets, while the associated accretion luminosity may provide an observable signature of ongoing formation \added[id=R4]{\citep{, keppler2018DiscoveryPlanetarymassCompanion, muller2018OrbitalAtmosphericCharacterization,
aoyama2018TheoreticalModelHydrogen, benisty2021CircumplanetaryDiskPDS70c, bowler2025VariabilityAurHubble, currie2025VLTMUSEDetection, stolker2025DirectImagingDiscovery}}.

\added[id=R1]{Quantifying the process of  gas accretion relies on multi-dimensional hydrodynamical simulation of planet-disk  interaction.}
Early studies employed two-dimensional (2D) hydrodynamical simulations, often introducing sink prescriptions to estimate gas accretion rates onto planets \citep{bryden2000InteractionProtoplanetsProtostellarDisks,nelson2000MigrationGrowthProtoplanetsa,tanigawa2002GasAccretionFlows}.
Because a planet carves a gap while strongly perturbing the flow on Hill-sphere scales, global simulations that simultaneously resolve both the disk-scale structure and the circumplanetary region are required to characterize gas accretion \citep{dangelo2002NestedgridCalculationsDiskplanet, dangelo2003OrbitalMigrationMass, bate2003ThreedimensionalCalculationsHighand}.
Moreover, the gap and the circumplanetary disk (CPD) are expected to be fed by the vertical flow, highlighting the importance of the vertical dimension for the mass supply and density distribution near the planet \citep{machida2008AngularMomentumAccretiona, machida2010GasAccretionProtoplanet, tanigawa2012DistributionAccretingGas, morbidelli2014MeridionalCirculationGas, szulagyi2014AccretionJupitermassPlanets}.
Three-dimensional (3D) global hydrodynamical simulations are therefore essential for a quantitative understanding of planetary gas accretion.

However, how gas accretion depends on disk thermodynamics remains poorly constrained.
Because high-resolution global 3D simulations are computationally expensive, many studies have adopted a locally isothermal equation of state, thereby avoiding the computational cost associated with radiative transfer \citep{choksi2023MaximumAccretionRate, li20233DGlobalSimulations, li2024ConcurrentAccretionMigration}.
Nonisothermal simulations with radiative cooling, however, suggest that locally isothermal models can overestimate accretion rates, although many of these calculations were limited to short-term integrations and thus may not have reached a quasi-steady state \citep{dangelo2003ThermohydrodynamicsCircumstellarDisks, ayliffe2009GasAccretionPlanetary, schulik2019Global3DRadiationhydrodynamic}.
In addition, radiative cooling can systematically modify gap profiles \citep{zhang2020EffectsDiscSelfgravitya, miranda2020PlanetDiskInteraction, ziampras2020ImportanceRadiativeEffects, zhang2024DependenceStructurePlanetopened}, indicating that thermodynamics can influence the global disk structure around the planet.

While the cooling dependence of accretion has been explored in local 3D or global 2D calculations \citep{zhu2016ShockdrivenAccretionCircumplanetary, wu2024EffectsThermodynamicsConcurrent}, systematic long-term global 3D surveys that reach a quasi-steady state remain scarce.
\added[id=R2]{Existing studies are often limited either to 2D, short integration times (typically tens to hundreds of orbits), or a restricted range of cooling times, making it difficult to isolate the thermodynamic dependence of accretion in a fully developed gap.}
Because accretion rates evolve alongside gap formation \citep{gressel2013GlobalHydromagneticSimulations,nelson2023GasAccretionJupitera}, measuring them in a quasi-steady state provides a more representative characterization than focusing solely on the initial transient.

In this work, we perform long-term 3D global hydrodynamical simulations, \added[id=R1]{typically for a timeframe of $10^3$ orbits,} to quantify gas accretion onto gap-opening planets.
Using a global grid combined with static mesh refinement, we resolve both the global disk structure and the Hill-sphere-scale flows.
We adopted a $\beta$-cooling prescription as a simplified model for radiative cooling, conducting a systematic survey over a wide range of cooling times to assess their impact on gas accretion rates.

The paper is structured as follows.
Section~\ref{sec:Numerical method} describes the numerical setup for our 3D hydrodynamical simulations.
In section \ref{sec:Numerical results}, we show that the gas accretion rate decreases with increasing cooling time.
Section~\ref{sec:Discussions} places our results in the context of previous work and discusses potential implications for planet growth and observations.
We summarize our findings in Sect.~\ref{sec:Conclusions}.

\begin{table*}[tp]
\caption{Parameters of hydrodynamical simulations.}
\resizebox{\textwidth}{!}{
\begin{tabular}{lccccccc}\hline\hline
     & $q=M_{\rm p}/M_\ast$ & $\beta=t_{\rm cool}\Omega_0$ & Levels of SMR  & $t_{\rm sink}\,[\Omega_0^{-1}]$ & $r_{\rm sink}\,[R_{\rm H}]$ & $r_{\rm sm}\,[R_{\rm H}]$ & $t_{\rm inj}\,[t_0]$ \\ \hline
     Fiducial runs       & $10^{-3},\,2\times10^{-3},\,3\times10^{-3}$ & $10^{-2},\,10^{-1},\,10^{0},\,10^{1},\,10^{2}$ & 2 & 0.2 & 0.1 & 0.1 & 1\\\hline
     Convergence tests   & $10^{-3}$ & $10^{-2},\,10^{0},\,10^2$ & 3 & 0.2 & 0.1 & 0.1 & 1 \\
                         & $10^{-3}$ & \added[id=R4]{$10^{0},\,10^2$} & 2 & 0.1 & 0.1 & 0.1 & 1 \\
                         & $10^{-3}$ & \added[id=R4]{$10^{0},\,10^2$} & 2 & 0.2 & 0.05 & 0.1 & 1 \\
                         & $10^{-3}$ & \added[id=R4]{$10^{0},\,10^2$} & 2 & 0.2 & 0.1 & 0.05 & 1 \\
                         & $10^{-3}$ & \added[id=R4]{$10^{0},\,10^2$} & 2 & 0.2 & 0.1 & 0.1 & 10 \\
     \hline
\end{tabular}
}
\tablefoot{\added[id=R2]{The following columns give the planet-star mass ratio, the dimensionless cooling time, the levels of static mesh refinement, the removal timescale, the sink radius, the smoothing length, and the injection time.}}
\label{tab:hydro simulations}
\end{table*}


\section{Numerical methods}\label{sec:Numerical method}
We simulated gas accretion onto a planet in a non-self-gravitating disk using the Athena++ code\footnote{We used the public version of Athena++ (ver.~24.0) which is available at: https://github.com/PrincetonUniversity/athena} \citep{stone2020AthenaAdaptiveMesh}. 
The simulations were performed in the spherical polar coordinates centered on the star, \added[id=R4]{$\bm{r}=(r,\,\theta,\,\phi)$}, where $r$ is the distance from the star, $\theta$ the polar angle, and $\phi$ the azimuthal angle. 
The planet is fixed on a circular orbit at $\bm{r}_{\rm p}=(r_0,\,\pi/2,\,0)$. 
We used the default numerical settings of Athena++, such as the integration schemes, unless otherwise specified.

\subsection{Governing equations}
We assumed that the gas is a compressible, viscous, and non-self-gravitating fluid.
The Athena++ code solves the following equations:
\begin{align}
    &\frac{\partial \rho}{\partial t}+\nabla\cdot(\rho\bm{v})=0,\\
    &\frac{\partial (\rho\bm{v})}{\partial t}+\nabla\cdot(\rho \bm{v}\bm{v}+P\bm{I}+\bm{\Pi})=-\rho\,\nabla\Phi,\\
    &\frac{\partial E}{\partial t}+\nabla\cdot\!\left[(E+P)\bm{v}+\bm{\Pi}\cdot\bm{v}\right]=-\rho\,\bm{v}\cdot\nabla\Phi+Q_{\rm cool}.\label{eq:energy equation}
\end{align}
Here $\rho$ is the density, $\bm{v}$ is the velocity, $P$ is the pressure, $\bm{I}$ is the identity tensor, and $\Phi$ is the gravitational potential. 
\added[id=R4]{The governing equations are solved in integral form using a finite-volume discretization.}

The total energy density is given by $E=e+\rho v^2/2$ and the internal energy density by $e=P/(\gamma-1)$, with $\gamma=1.4$ being the adiabatic index.
A thermal relaxation term, $Q_{\rm cool}$, was included on the right-hand side of Eq.~\ref{eq:energy equation} and is described in Sect.~\ref{sec:Cooling term}.

\added[id=R4]{The viscous stress tensor, including only shear viscosity and neglecting bulk viscosity, is defined as}
\begin{align}
    \Pi_{ij}=\rho\nu\Bigg(\frac{\partial v_i}{\partial x_j}+\frac{\partial v_j}{\partial x_i}-\frac{2}{3}\delta_{ij}\nabla\cdot\bm{v}\Bigg),
\end{align}
where $\nu=\alpha c_{\rm s}H$ is the kinematic viscosity, $\alpha$ the Shakura-Sunyaev dimensionless viscous parameter \citep{shakura1973BlackHolesBinary}, and $c_{\rm s}$ the sound speed.

The gravitational potential includes contributions from the stellar gravity, the planetary gravity, and the indirect term \added[id=R4]{arising from the acceleration of the star-centered frame by the planet,
\begin{align}
    \Phi=-\frac{GM_\ast}{|\bm{r}|}-\frac{GM_{\rm p}}{(|\bm{r}_{\rm p}-\bm{r}|^2+\epsilon^2)^{1/2}}+\frac{GM_{\rm p}}{|\bm{r}_{\rm p}|^3}\bm{r}_{\rm p}\cdot\bm{r}.\label{eq:grav potential}
\end{align}
Here, $G$ is the gravitational constant, $M_\ast$ is the stellar mass, $M_{\rm p}$ is the planet mass,} and $\epsilon$ is the smoothing length. 
The smoothing length was set to $\epsilon = 0.1\,R_{\rm H}$ \citep{li20233DGlobalSimulations}, \added[id=R4]{where $R_{\rm H} = r_0 (M_{\rm p}/(3M_\ast))^{1/3}$ is the Hill radius}.
Hereafter, we define \added[id=R4]{$r_{\rm cyl} \equiv |\bm{r}-\bm{r}_{\rm p}|$}.
\added[id=R4]{We note that Eq.~\ref{eq:grav potential} includes only the planet-induced indirect term. 
More general treatments that also include disk-induced indirect terms have recently been proposed for nonaxisymmetric disk dynamics  \citep{crida2025ReflexInstabilityExponential,crida2025InertialForcesIndirect}.}
The gravity of the planet was gradually inserted into the disk to prevent shock formation. 
Following \cite{kanagawa2023KinematicSignaturesLowmass}, we used the following injection function:
\begin{empheq}
    [left={M_{\rm p}(t)=\empheqlbrace}]{alignat=2}
    &M_{\rm p}\sin^2\Bigg(\frac{\pi}{2}\frac{t}{t_{\rm inj}}\Bigg)&&\quad\text{when}\,t<t_{\rm inj},\\
    &M_{\rm p}&&\quad\text{when}\,t\geq t_{\rm inj},
\end{empheq}
where $t$ is the time, $t_{\rm inj}=t_0\equiv2\pi/\Omega_0$ is the injection time, and $\Omega_0=\sqrt{GM_\ast/r_0^3}$ is the orbital frequency at $r=r_0$. 

\subsection{Cooling term}\label{sec:Cooling term}
We treated the cooling of the gas as a thermal relaxation process \citep[e.g.,][]{gammie2001NonlinearOutcomeGravitational}. This was implemented as a source term of the energy equation,
\begin{align}
Q_{\rm cool}=-\frac{e-e_{0}}{t_{\rm cool}},\label{eq:q cool}
\end{align}
where $e_0$ represents the initial value. The isothermal and adiabatic limits are achieved when the cooling timescale approaches, respectively, $t_{\rm cool}\rightarrow0$ and $t_{\rm cool}\rightarrow\infty$.

\subsection{Sink cell}\label{sec:Sink cell}
Gas accretion onto the planet is modeled using a sink cell prescription \citep[e.g.,][]{federrath2010ModelingCollapseAccretion}.
Following \citet{li2021AccretionGasGiants, li20233DGlobalSimulations}, the following prescription was adopted:
\begin{align}
    \frac{\partial X}{\partial t}\bigg|_{r_{\rm cyl}\leq r_{\rm sink}}=-\frac{X}{t_{\rm sink}}.
\end{align}
The sink removes mass, momentum, and energy proportionally, with $X\in\{\rho,\,\rho\bm{v},\,E\}$.
\added[id=R4]{Following \cite{li20233DGlobalSimulations}, the sink radius and timescale are set to $r_{\rm sink}=0.1\,R_{\rm H}$ and $t_{\rm sink}=0.2\,\Omega_0^{-1}$, respectively.
Appendix~\ref{sec:Convergence tests} shows that the measured accretion rate depends only weakly on $t_{\rm sink}$ and, except in the slow-cooling regime, on $r_{\rm sink}$.
The removed gas is not added to the planet, and the planet mass is kept fixed to isolate the dependence of the accretion rate on the cooling time at fixed planet mass.}
The gas accretion rate at each time step is computed by
\begin{align}
    \dot{M}_{\rm acc}=2\int_{r_{\rm cyl}\leq r_{\rm sink}}\frac{\rho}{t_{\rm sink}}\mathrm{d}V,
\end{align}
where the factor of 2 accounts for the assumed symmetry with respect to the disk midplane (see Sect.~\ref{sec:Resolutions and boundary conditions}).

\subsection{Initial disk model}
The initial disk model adopted a radial temperature profile $T \propto r^\xi$, with $\xi = -0.5$.
The initial density profile follows \citep{nelson2013LinearNonlinearEvolution},
\begin{align}
    \rho(R,z)=\rho_0\left(\frac{R}{r_0}\right)^{\chi}\!\exp\!\left[\frac{GM_\ast}{c_{\rm s}^2}\!\left(\frac{1}{\sqrt{R^2+z^2}}-\frac{1}{R}\right)\right],
\end{align}
where $R = r\sin\theta$ and $z = r\cos\theta$, and \added[id=R4]{$\chi = -2.25$}. 
This choice yields a surface density profile $\Sigma \propto r^{-1}$.
The azimuthal velocity is given by $v_\phi = R\,\Omega(R,z)$, \added[id=R4]{with
\begin{align}
    \Omega(R,z)=\Omega_{\rm K}(R)\left[1+\frac{h^2}{2}\left(\chi+\xi+\frac{\xi}{2}\frac{z^2}{H^2}\right)\right],
\end{align}
\citep{takeuchi2002RadialFlowDust}. Here} $\Omega_{\rm K}(R)=\sqrt{GM_\ast/R^3}$ is the Keplerian angular frequency, $H = c_{\rm s}/\Omega_{\rm K}$ is the pressure scale height, and $h \equiv H/R$.
The disk aspect ratio at the planet location was set to $h_0=H_0/r_0=0.05$
\added[id=R4]{\citep[e.g.,][]{isella2016RingedStructuresHD}.}
The initial radial and polar velocities were set to zero, $v_r = v_\theta = 0$.

\subsection{Resolutions and boundary conditions}\label{sec:Resolutions and boundary conditions}
We used a uniformly spaced radial grid from $r_{\rm in}=0.4$ to $r_{\rm out}=2.4$ with 256 cells; the polar and azimuthal angles are resolved by 16 and 768 cells, respectively. 
Owing to a symmetry, we simulated only the upper half of the disk, up to four scale heights at $r=r_0$, \added[id=R4]{namely, $\theta\in[\pi/2-\arctan(4h_0),\pi/2]$}. 
The full azimuthal domain is covered, $\phi\in[-\pi,\pi]$. 
We applied \added[id=R2]{two} levels of static mesh refinement (SMR) inside the Hill sphere, achieving a resolution of approximately \added[id=R2]{30--44 cells} per dimension across the Hill sphere, depending on the planet mass (Sect.~\ref{sec:Code units and simulation parameters}).
Appendix~\ref{sec:Convergence tests} provides the convergence tests.

In the polar and azimuth directions, we adopted reflecting and periodic boundary conditions. 
A fixed boundary condition was imposed in the radial direction, in which the density and the velocity are set to the initial values. 
We used the wave-damping zone, in which an arbitrary physical quantity is damped by the following rates \citep{deval-borro2006ComparativeStudyDisc, choksi2023MaximumAccretionRate}:
\begin{align}
    \frac{\partial Y}{\partial t}=-\frac{Y-Y_0}{t_{\rm damp}(r)}\,f_{\rm damp}(r),
\end{align}
where 
\begin{empheq}
    [left={f_{\rm damp}(r)=\empheqlbrace}]{alignat=2}
    &\Bigg[1-\sin^2\bigg(\frac{1}{2}\frac{r-r_{\rm in}}{r_{\rm damp,in}-r_{\rm in}}\bigg)\Bigg]&&\quad\text{at}\,r\leq r_{\rm damp,in},\\
    &\sin^2\bigg(\frac{1}{2}\frac{r-r_{\rm damp,out}}{r_{\rm out}-r_{\rm damp,out}}\bigg)&&\quad\text{at}\,r\geq r_{\rm damp,out}.
\end{empheq}
Here we set $r_{\rm damp,in}=1.25\,r_{\rm in}$, $r_{\rm damp,out}=0.75\,r_{\rm out}$, and $t_{\rm damp}=2\pi/\Omega_{\rm K}(r)$ ($r=r_{\rm damp,in}$ or $r_{\rm damp,out}$). Following \cite{li20233DGlobalSimulations}, we only damped \added[id=R2]{the radial velocity} and left other quantities undamped

\begin{figure}[tp]
    \centering
    \includegraphics[width=1\linewidth]{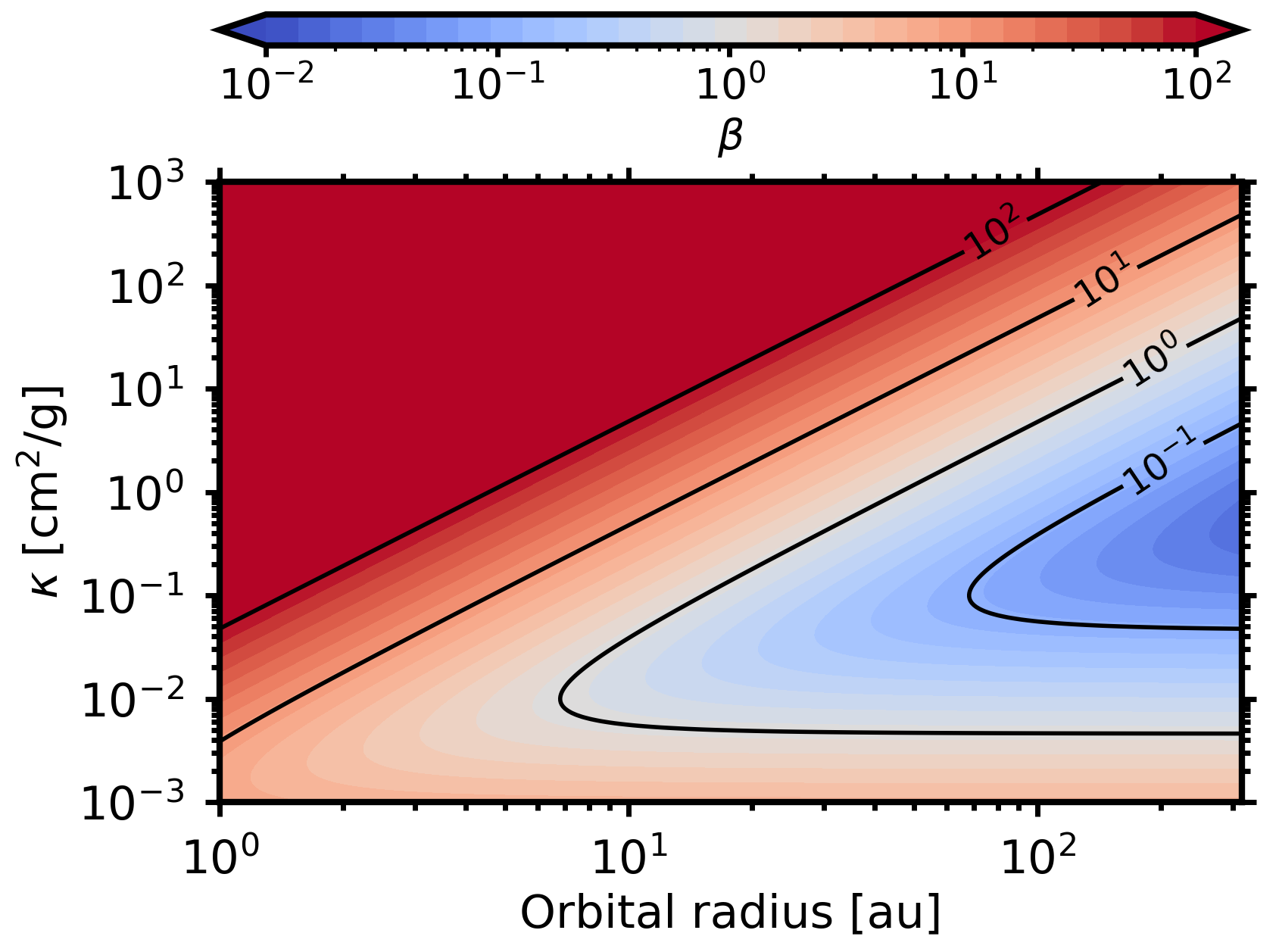}
    \caption{Dimensionless cooling time of the background disk as a function of opacity and orbital radius. The values of $\beta$ are computed using Eq.~\ref{eq:beta disk} (Appendix~\ref{sec:Cooling time estimation}), assuming $\Sigma = 10^3\,\mathrm{g\,cm^{-2}}(r/1\,\mathrm{au})^{-1}$ and $T = 150\,\mathrm{K}(r/1\,\mathrm{au})^{-0.5}$.}
    \label{fig:cooling_time_contour}
\end{figure}

\begin{figure*}[tp]
    \centering
    \includegraphics[width=1\linewidth]{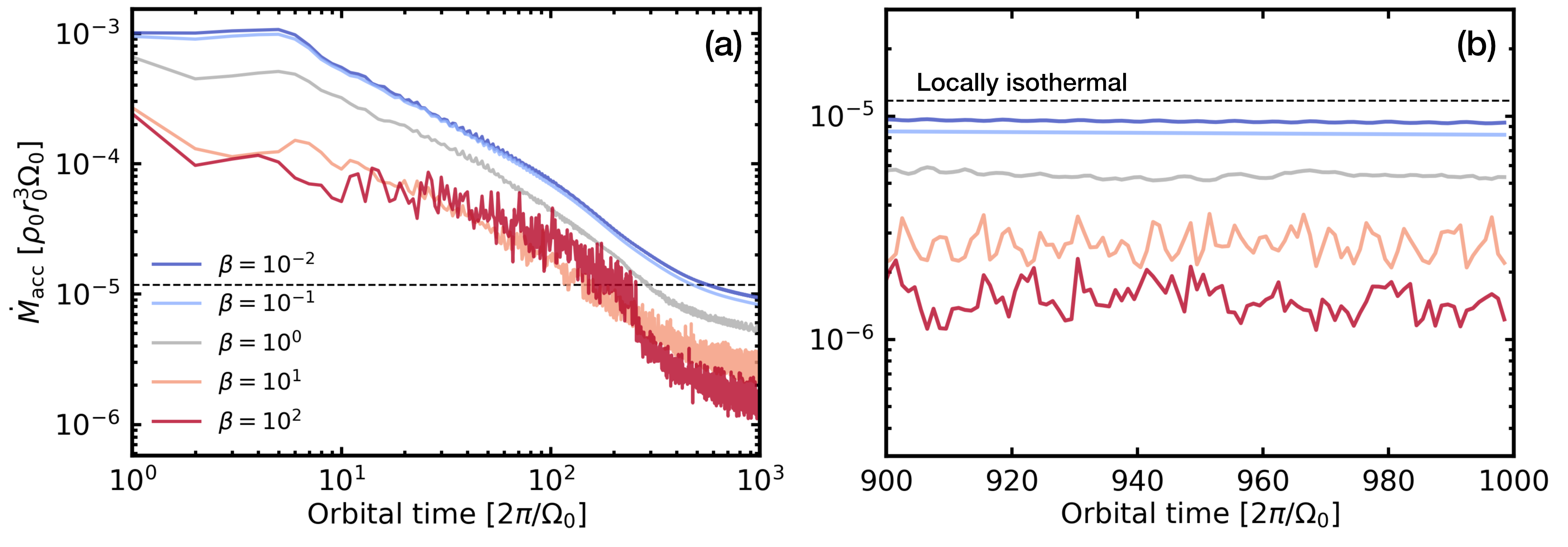}
    \caption{Time evolution of the gas accretion rate onto a planet with $q=10^{-3}$ for different cooling times, $\beta$. The horizontal dashed line shows the prediction of a locally isothermal model (Eq.~\ref{eq:mdot locally iso}). \textit{Left:} Full time series, with $\dot{M}_{\rm acc}$ measured once per orbit. \textit{Right:} Zoom into the last 100 orbits.}
    \label{fig:1d_plot_mdot}
\end{figure*}

\begin{figure}[tp]
    \centering
    \includegraphics[width=1\linewidth]{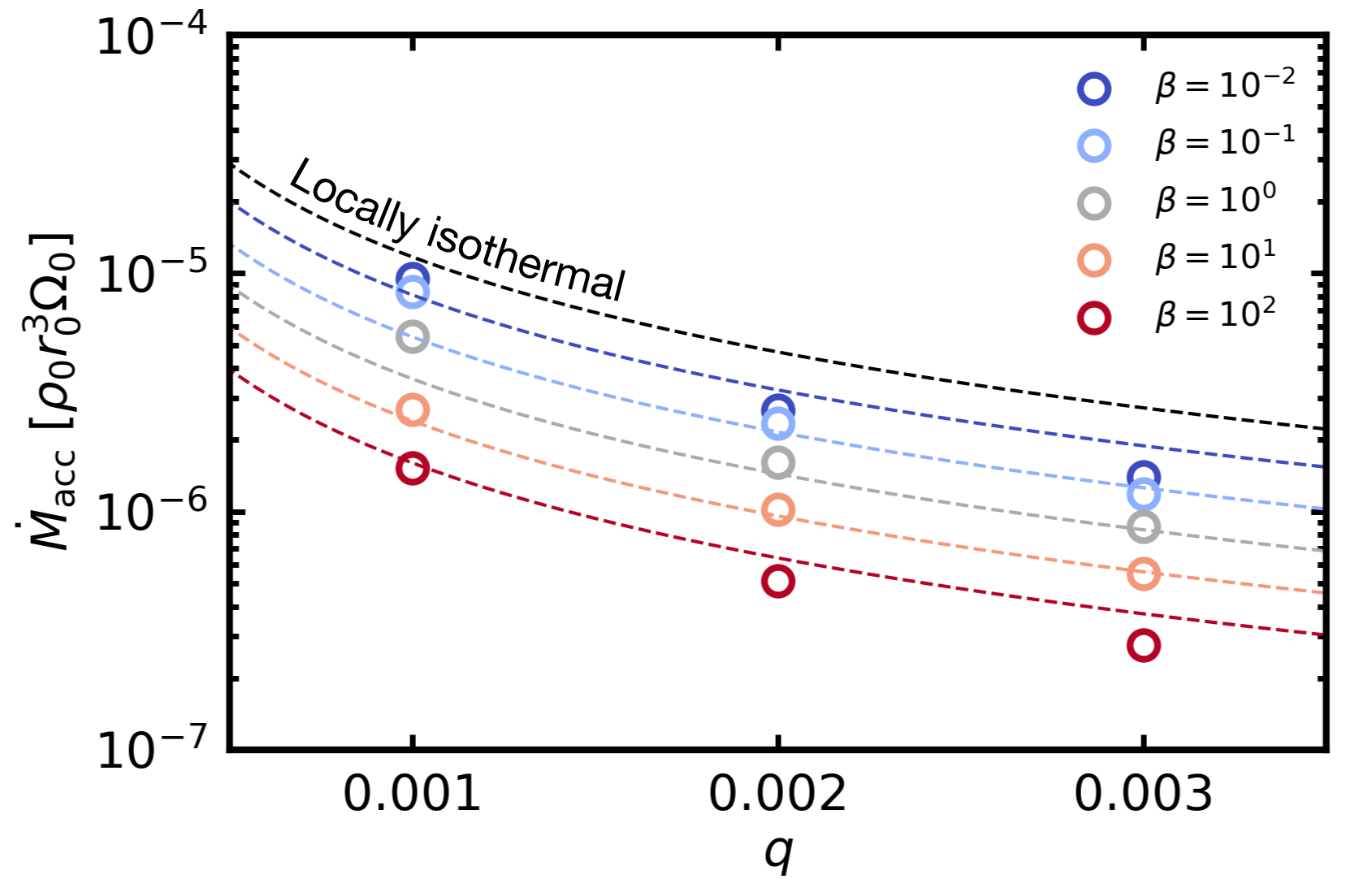}
    \caption{Quasi-steady gas accretion rates as functions of the planet–star mass ratio, $q$, and the dimensionless cooling time, $\beta$. 
    Values are obtained by averaging over the last \added[id=R2]{100} orbits. 
    The black dashed curve shows the locally isothermal model (Eq.~\ref{eq:mdot locally iso}). Colored \added[id=R2]{dotted} curves show the empirical formula given by Eq.~\ref{eq:mdot empirical}.}
    \label{fig:1d_mdot_vs_qpl}
\end{figure}

\subsection{Code units and simulation parameters}\label{sec:Code units and simulation parameters}
Our simulations were performed in the units of $G=M_\ast=\rho_0=r_0=\Omega_0=1$. 
We evolved the system for $10^3$ planetary orbits, sufficient to reach a quasi-steady state in the gap region, using the second-order orbital advection scheme (\texttt{OAorder}=2). 
\added[id=R4]{We assumed a fixed effective viscosity, $\alpha=10^{-3}$, without specifying a particular turbulent mechanism.
This value allows a quasi-steady gap to form within the duration of our 3D simulations, while avoiding the extremely low-viscosity regime where planet-induced gap edges can become vortex dominated \citep[e.g.,][]{fu2014LongtermEvolutionPlanetinduced, hammer2017SlowlygrowingGapopeningPlanets}.
}

We parameterized the planet–star mass ratio and the dimensionless cooling time as
\begin{align}
q\equiv \frac{M_{\rm p}}{M_\ast},\quad \beta\equiv t_{\rm cool}\,\Omega_0.
\end{align}
We considered \added[id=R1]{$q=10^{-3},\,2\times10^{-3}$, and $3\times10^{-3}$}, corresponding to approximately 1--3 Jupiter masses around a solar-mass star.
\added[id=R1]{A cooling timescale between $\beta=10^{-2}$ and $\beta=10^{2}$} was explored.
For simplicity, we adopted a constant $\beta$ throughout the computational domain. The chosen range of $\beta$ covers a wide range of disk conditions (Fig.~\ref{fig:cooling_time_contour}).
\added[id=R2]{We confirmed that the numerical timestep is always $\lesssim 3\times10^{-3}\,\Omega_0^{-1}$, ensuring that the cooling term is properly resolved even for the shortest cooling time, $t_{\rm cool}=10^{-2}\,\Omega_0^{-1}$ (Eq.~\ref{eq:q cool}).}


\begin{figure*}[tp]
    \centering
    \includegraphics[width=0.99\linewidth]{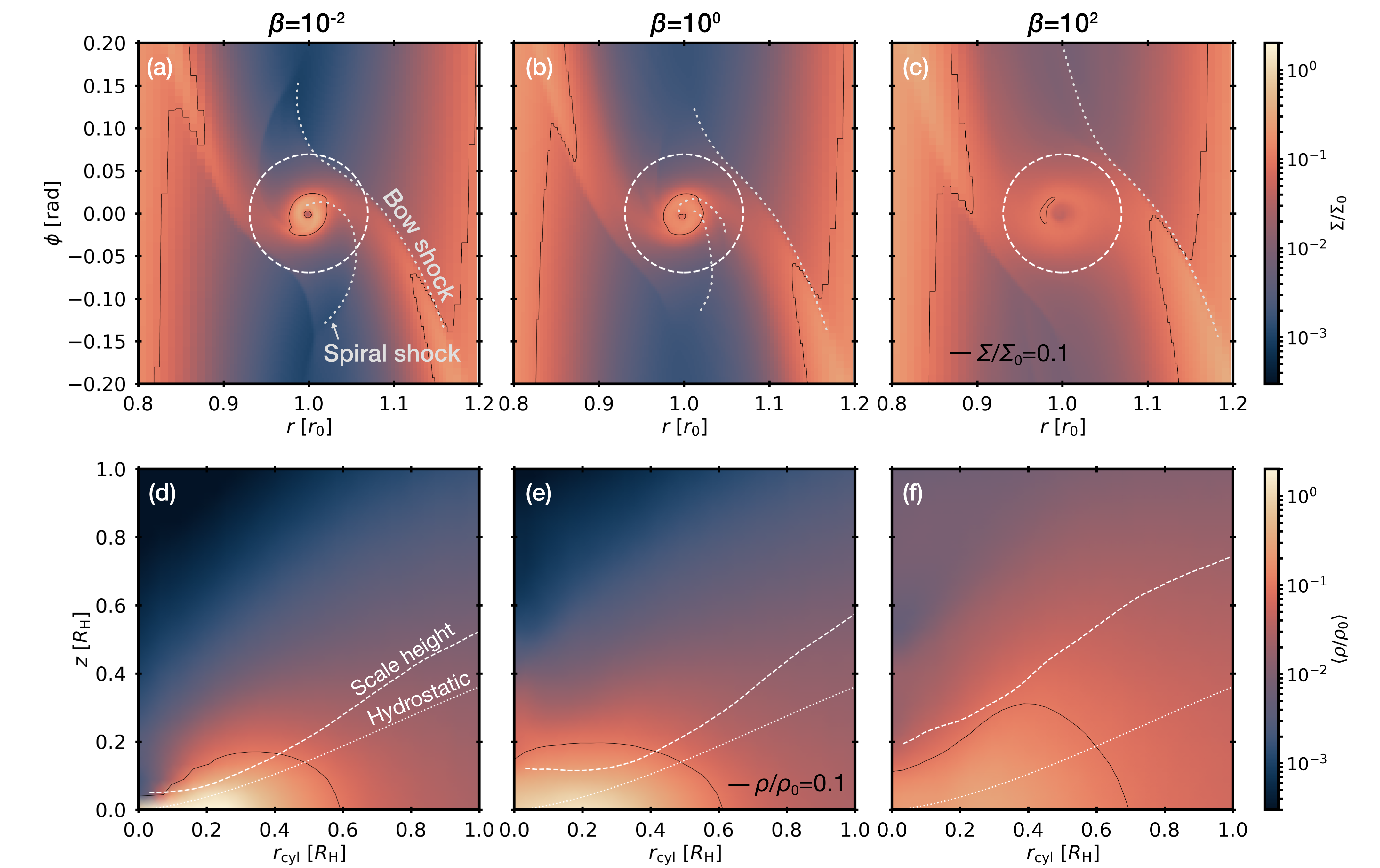}
    \caption{Surface density \added[id=R2]{(\textit{top})} and azimuthally averaged volume density in the circumplanetary disk region \added[id=R2]{(\textit{bottom})} for different cooling times, $\beta$. The planet–star mass ratio is $q=10^{-3}$. \added[id=R1]{All panels are snapshots at the end of the simulation, $t=10^3\,t_0$.} \added[id=R4]{The black curves indicate the isocontours of $\Sigma/\Sigma_0 = 0.1$ (\textit{top}) and $\rho/\rho_0=0.1$ (\textit{bottom}), respectively.} \textit{Top:} \added[id=R2]{Dotted curves qualitatively indicate the bow shock and the tidally induced spiral shock, inferred from the divergence of the velocity field (Fig.~\ref{fig:2d_slice_local_div_v}). For clarity, they are shown only on the right-hand side of each panel.} \textit{Bottom:} Vertical slice. \added[id=R1]{The dashed and dotted curves indicate the numerically obtained scale height and the analytic estimate based on vertical hydrostatic equilibrium in an isothermal CPD (Eq.~\ref{eq:cpd scale height}), respectively.}}
    \label{fig:2d_slice_local_sigma_rho}
\end{figure*}

\begin{figure}[tp]
    \centering
    \includegraphics[width=1\linewidth]{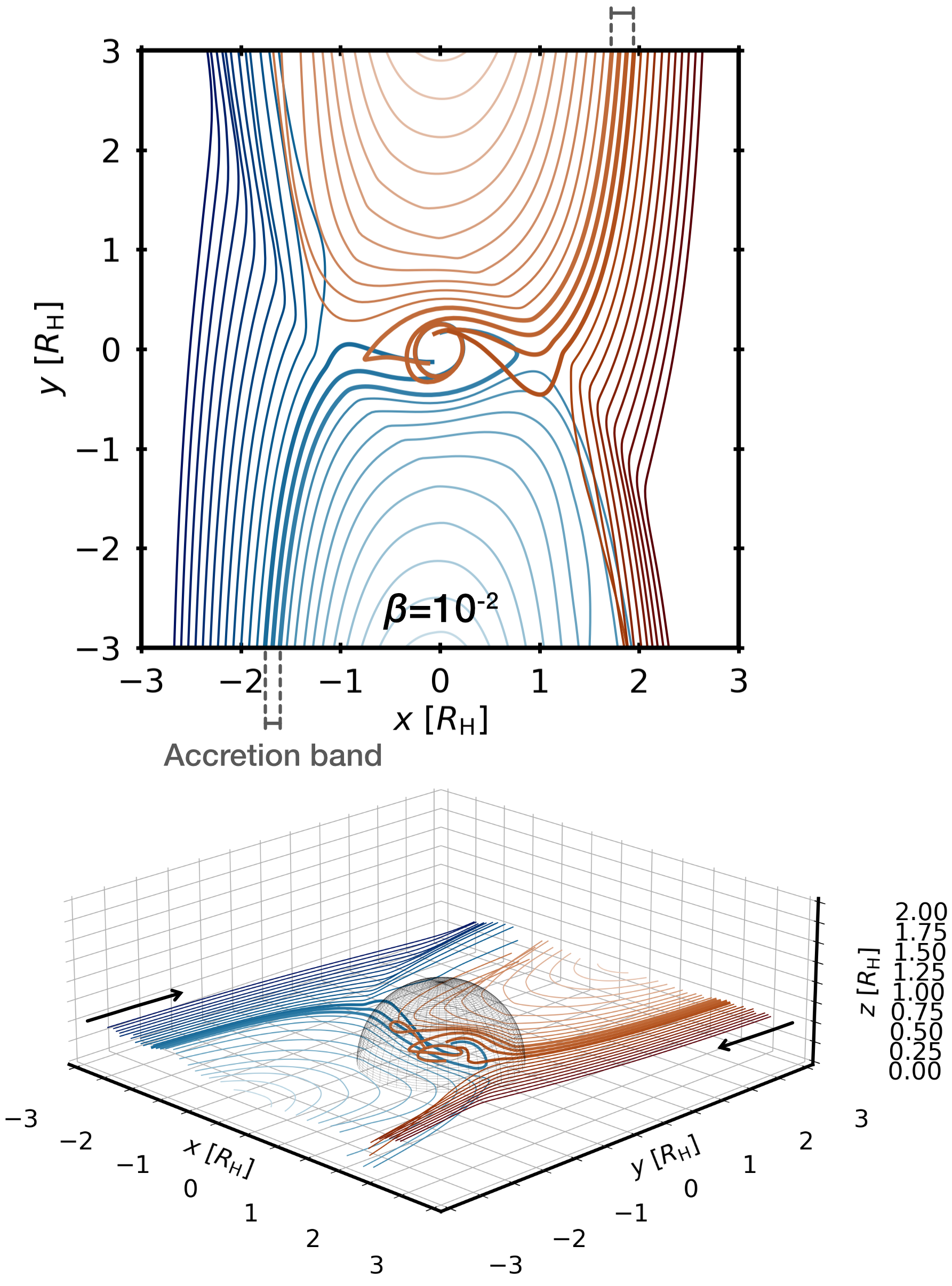}
    \caption{\added[id=R1]{Streamlines launched from $z_0 = 0.5\,R_{\rm H}$ at the end of the simulation.} We set $q=10^{-3}$ and $\beta=10^{-2}$.
    \added[id=R1]{The thick curve traces a representative streamline of gas that passes through the accretion bands marked by dashed lines and reaches radii $r<0.2\,R_{\rm H}$. The sphere in the bottom panel indicates the Hill region.}
    }
    \label{fig:3d_streamlines_b1e-02}
\end{figure}

\begin{figure}[tp]
    \centering
    \includegraphics[width=1\linewidth]{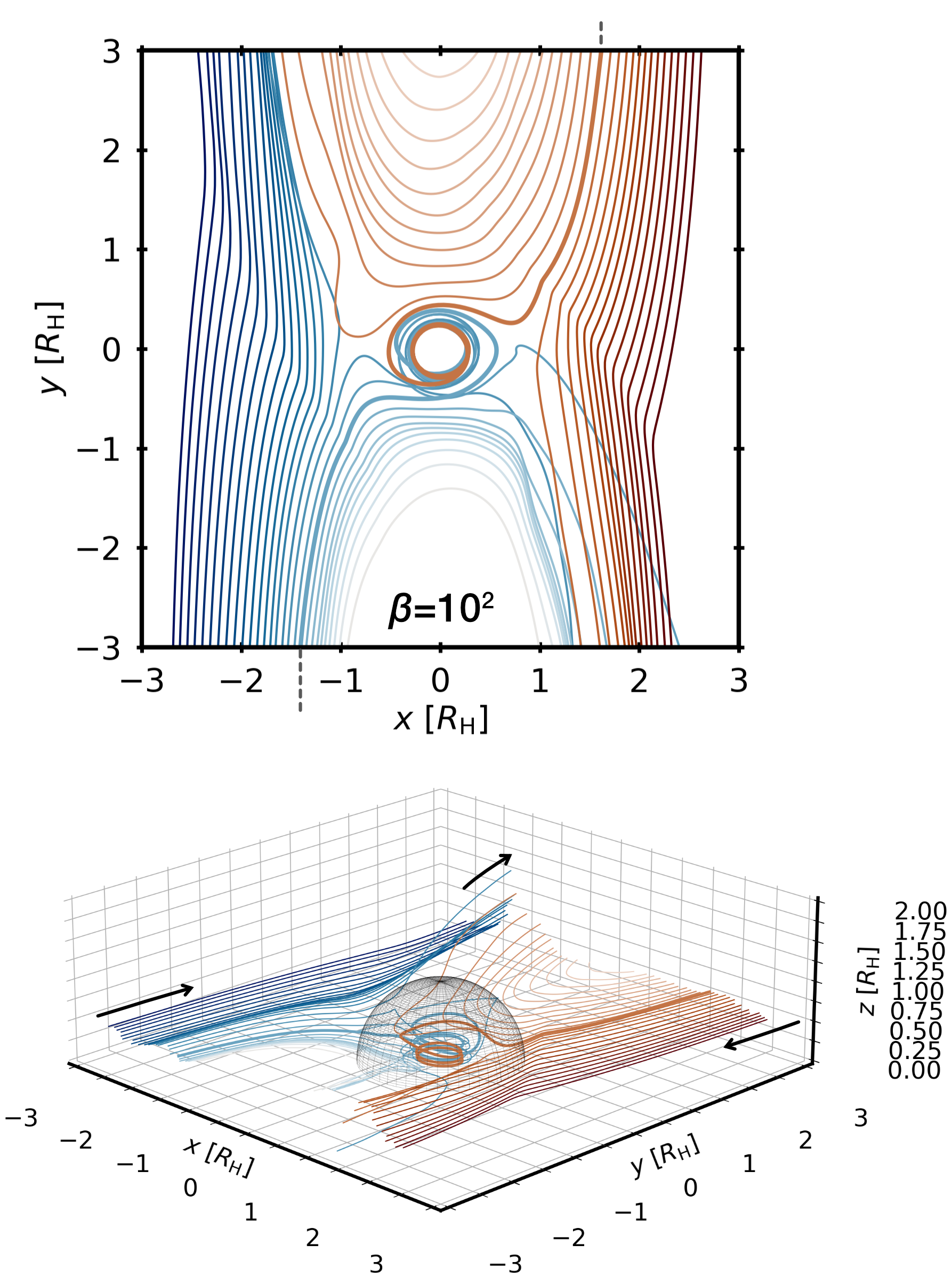}
    \caption{Same as Fig.~\ref{fig:3d_streamlines_b1e-02}, but for $\beta=10^2$.}
    \label{fig:3d_streamlines_b1e+02}
\end{figure}

\begin{figure}[tp]
    \centering
    \includegraphics[width=1\linewidth]{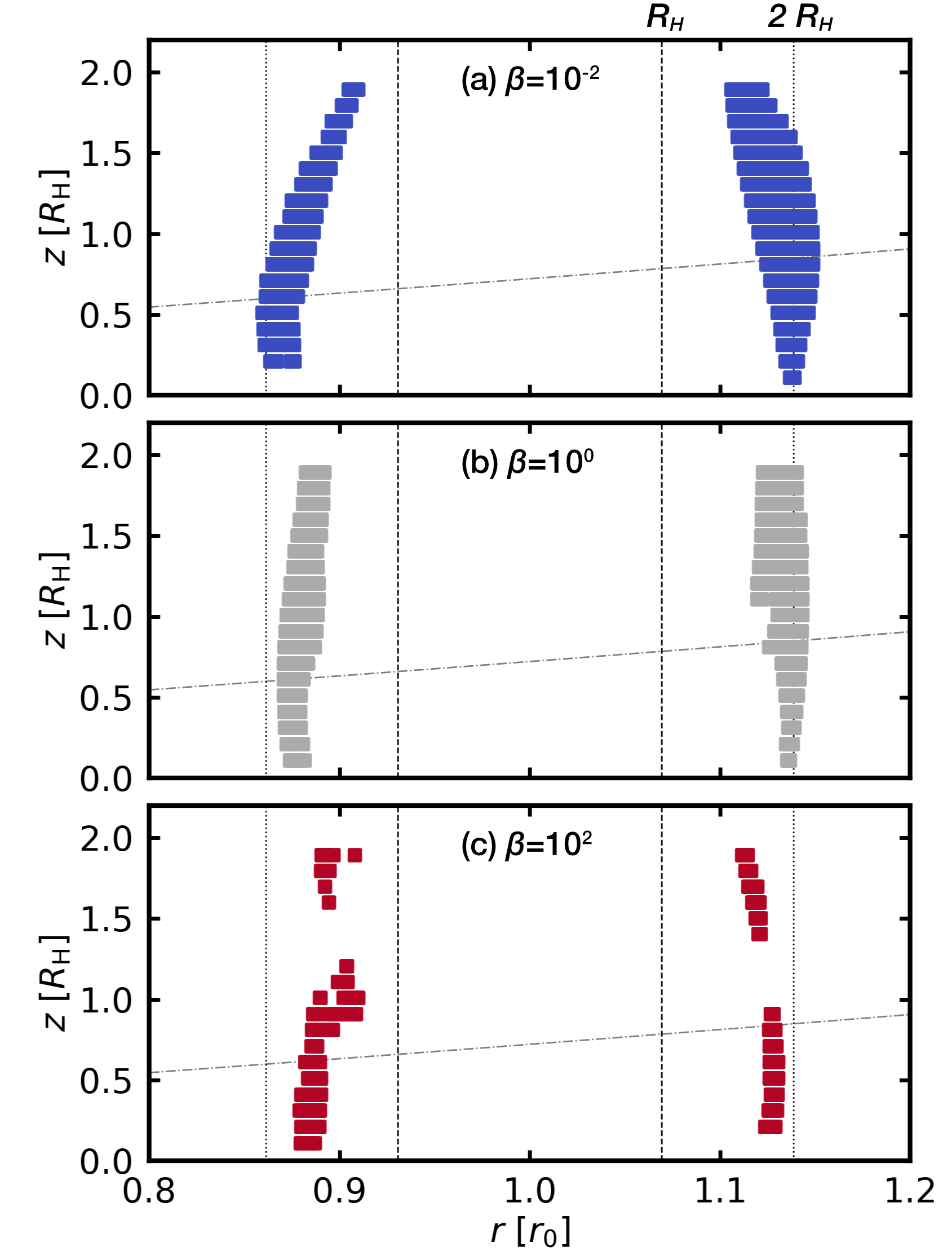}
    \caption{Vertical structure of the accretion band for different cooling times, $\beta$. We set $q=10^{-3}$. We integrate streamlines only within the region $z \leq 2\,R_{\rm H}$ and $\phi \in [-12\,H_0/r_0,\,12\,H_0/r_0]$, and stop the integrations when they reach $r_{\rm cyl} < 0.2,R_{\rm H}$ \citep[cf.][]{maeda2022DeliveryGasCircumplanetary}. \added[id=R1]{The dot-dashed line marks the gas scale height, $H$.}}
    \label{fig:accretion_bands}
\end{figure}

\begin{figure}[tp]
    \centering
    \includegraphics[width=1\linewidth]{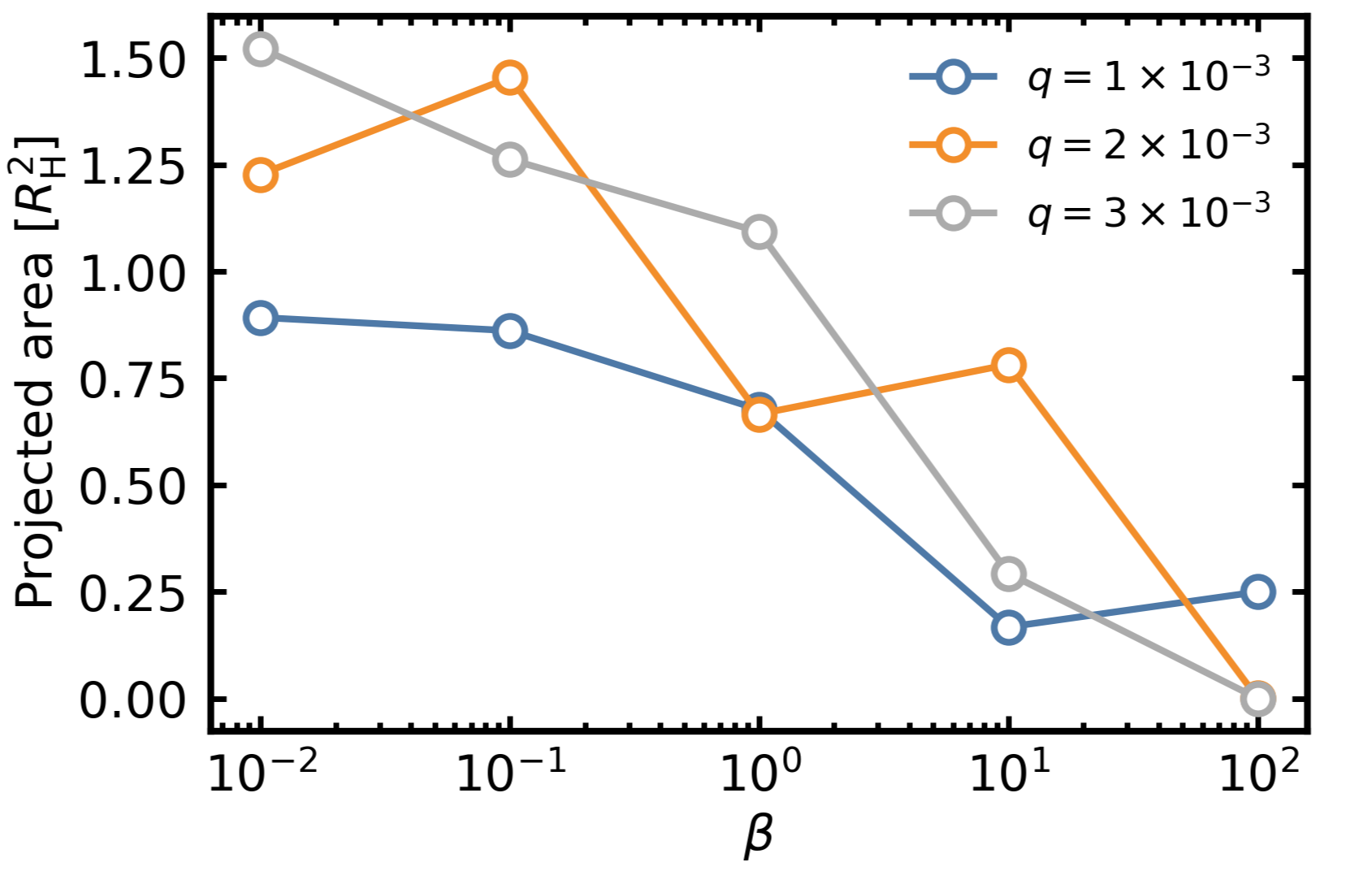}
    \caption{\added[id=R4]{Total projected area of the inner and outer accretion bands in the $r$–$z$ plane} as a function of $\beta$, for different planet-star mass ratios, $q$.}
    \label{fig:accretion_cross_section}
\end{figure}

\section{Numerical results}\label{sec:Numerical results}
We find that the gas accretion rate onto a gap-opening planet decreases with increasing cooling time, $\beta$, largely independent of the planet–star mass ratio, $q$ (Sect.~\ref{sec:Gas accretion rates}).
To identify what controls this $\beta$-dependence, we next examine the thermal and density structures of the circumplanetary region (Sect.~\ref{sec:Thermal and density structure of the circumplanetary region}) and analyze the gas streamlines accreting onto the circumplanetary region (Sect.~\ref{sec:Cooling-regulated mass supply to the CPD}).

\subsection{Gas accretion rates}\label{sec:Gas accretion rates}
Inefficient gas cooling leads to a reduced gas accretion rate onto gap-opening planets.
Figure~\ref{fig:1d_plot_mdot} shows the time evolution of the gas accretion rate onto the planet with $q=10^{-3}$.
The gas accretion rate decreases with time as the planet opens a gap in the disk, and then reaches a quasi-steady state.
\added[id=R2]{When $\beta\leq10^{-1}$}, the quasi-steady state value is comparable to the value predicted by a locally isothermal model \added[id=R2]{within an error of 30\%}.
The quasi-steady state value decreases as $\beta$ increases. 
When $\beta=10^2$, $\dot{M}_{\rm acc}$ is an order of magnitude lower than the locally isothermal prediction.

\added[id=R4]{The time required to reach a quasi-steady accretion state is expected to be closely related to the gap-opening timescale, which can be estimated as \citep{zhang2024DependenceStructurePlanetopened}
\begin{align}
    t_{\rm gap}=\frac{r_0^2}{\nu}\Bigg(\frac{w_{\rm gap}}{2r_0}\Bigg)^2\simeq1.34\times10^3\,\Bigg(\frac{10^{-3}}{\alpha}\Bigg)\,t_0,\label{eq:t_gap}
\end{align}
where $w_{\rm gap}/r_0\simeq5.8\,h_0$ \citep{dong2017WhatMassGapopening}. 
Since this timescale is comparable to our integration time, we additionally performed longer integrations to assess the convergence of the measured accretion rates (Appendix~\ref{sec:Convergence tests}).
The longer integrations confirm that our fiducial integration time is sufficient to capture the quasi-steady accretion behavior.
}

Figure~\ref{fig:1d_mdot_vs_qpl} summarizes the quasi-steady gas accretion rate as a function of $\beta$ and $q$.
Over the parameter range explored in this study, the $\beta$-dependence of the gas accretion rate shows little dependence on $q$.
\added[id=R4]{In Section~\ref{sec:Empirical formula for gas accretion rates}, we derive an empirical formula for the gas accretion rate, depicted as dashed curves in Fig.~\ref{fig:1d_mdot_vs_qpl}}.

In the following sections, we examine the circumplanetary region in more detail to identify the physical origin of this $\beta$-dependence.

\subsection{Thermal and density structure of the circumplanetary region}\label{sec:Thermal and density structure of the circumplanetary region}

Inefficient cooling modifies the density and vertical structure in the circumplanetary region, as illustrated in Fig.~\ref{fig:2d_slice_local_sigma_rho}, which shows the gas surface density (top panels) and vertical slices of the gas volume density (bottom panels).
For $\beta \lesssim 10^0$, a disk-like structure is present within the Hill region.
When $\beta=10^2$, the gas within the Hill region is less dense and vertically extended, resembling a pressure-supported, envelope-like structure (Fig.~\ref{fig:2d_slice_local_sigma_rho}f).
Hereafter, we use the term CPD to denote the gas within the Hill region, irrespective of whether it exhibits a disk-like or envelope-like structure.

A bow shock exterior to the planet’s orbit and tidally induced spiral shocks form around the CPD. 
The gas is heated at these shock fronts (Fig.~\ref{fig:2d_slice_temperature}).
As $\beta$ increases, the bow shock moves farther away from the planet, and the spiral shock becomes progressively indistinguishable.

As $\beta$ increases, the temperature within the Hill region rises, leading to an increase in the gas scale height (Figs.~\ref{fig:2d_slice_local_sigma_rho}d--f). 
In fast cooling regime, \added[id=R2]{$\beta=10^{-2}$}, the scale height closely follows the analytic prediction for an isothermal CPD at approximately \added[id=R4]{$r_{\rm cyl}<\,R_{\rm H}/3$} \added[id=R1]{\citep{quillen1998ProtojovianPlanetsDrive}}.
For larger $\beta$ values, the scale heights increase by a factor of 2--3.
The gas surface density within the Hill region, in particular at $r_{\rm cyl} < 0.3\,R_{\rm H}$, decreases with increasing $\beta$ (Fig.~\ref{fig:2d_slice_local_sigma_rho}a--c).
As a consequence, the volume density in the CPD decreases with increasing $\beta$.
In the next section, we analyze the gas streamlines accreting onto the CPD to assess how the mass supply to the CPD depends on the cooling time.

\subsection{Cooling-regulated mass supply to the CPD}\label{sec:Cooling-regulated mass supply to the CPD}
The mass supply onto the CPD becomes less efficient as the cooling time increases.
Figure~\ref{fig:3d_streamlines_b1e-02} shows gas streamlines around the circumplanetary region \added[id=R1]{for $\beta=10^{-2}$}, launched from a height of $z_0 = 0.5\,R_{\rm H}$, \added[id=R1]{while Fig.~\ref{fig:3d_streamlines_b1e+02} for $\beta=10^2$.}
The CPD is predominantly supplied by gas flows passing through a narrow region, hereafter referred to as the accretion band indicated by dashed lines in Figs.~\ref{fig:3d_streamlines_b1e-02} and \ref{fig:3d_streamlines_b1e+02}, which is consistent with previous findings \citep[e.g.,][]{tanigawa2012DistributionAccretingGas,maeda2022DeliveryGasCircumplanetary}.
\added[id=R1]{We note that gas motion along the horseshoe streamline differs substantially between the two $\beta$ cases (bottom panels of Figs.~\ref{fig:3d_streamlines_b1e-02} and \ref{fig:3d_streamlines_b1e+02}); this difference is discussed further in Sect.~\ref{sec:Discussions}.
}

We find that the \added[id=R4]{projected area of the accretion bands in the $r$–$z$ plane decreases} with increasing $\beta$.
Figure~\ref{fig:accretion_bands} shows the vertical profile of the accretion bands for different cooling times. 
The accretion bands are located at approximately \added[id=R4]{$r \simeq r_0 \pm 2\,R_{\rm H}$} and extend vertically over $z>R_{\rm H}$.
Figure~\ref{fig:accretion_cross_section} further shows \added[id=R4]{total projected area of the inner and outer accretion bands in the $r$–$z$ plane} as a function of $\beta$ for different planet-star mass ratios. 
\added[id=R4]{This projected area} decreases as $\beta$ increases independent of $q$.
These results suggest that inefficient cooling leads to weaker shocks, as higher gas temperatures reduce the characteristic Mach numbers of the flow. 
Energy dissipation at the shocks regulates the cross section of the accretion band \citep{tanigawa2002GasAccretionFlows}.
Consequently, the mass supply into the CPD becomes inefficient for larger $\beta$, which leads to a lower gas surface density in the CPD observed in Fig.~\ref{fig:2d_slice_local_sigma_rho}.
As $\beta$ increases, the gas surface density in the CPD decreases while the scale height increases, leading to a lower volume density.
As a result, the gas accretion rate onto the planet decreases with increasing $\beta$ (Fig.~\ref{fig:1d_mdot_vs_qpl}).

\section{Discussions}\label{sec:Discussions}

\subsection{Empirical formula for gas accretion rates}\label{sec:Empirical formula for gas accretion rates}
Here we introduce an empirical formula for the gas accretion rate onto gap-opening planets.
Under locally isothermal conditions, gas accretion can be described by the Hill accretion rate.
Following \cite{choksi2023MaximumAccretionRate} and \cite{li20233DGlobalSimulations}, \added[id=R4]{for the superthermal mass limit, we define
\begin{align}
    \dot{M}_{\rm acc}^{\rm Liso}\!(q)\equiv\pi R_{\rm H}^2h_0\rho_{\rm p}r_0\Omega_0, \label{eq:mdot locally iso}
\end{align}
where} $\rho_{\rm p}$ denotes \added[id=R2]{the midplane gas density at the planet's Hill radius.}
\cite{li20233DGlobalSimulations} estimated this density as \added[id=R4]{(their Eq. 16)}
\begin{align}
    \rho_{\rm p}=\frac{\Sigma_{\rm gap}}{\sqrt{2\pi}H_{\rm CPD}\!(R_{\rm H})},\label{eq:rho p}
\end{align}
\added[id=R4]{with 
\begin{align}
    \Sigma_{\rm gap}=\frac{\Sigma_0}{1+0.04\,K},\quad K=q^2h_0^{-5}\alpha^{-1},\quad\Sigma_0=\sqrt{2\pi}\rho_0H_0\label{eq:Kanagawa gap}
\end{align}
being} the surface density at the bottom of the gap \citep{kanagawa2015MassEstimatesGiant}.
Assuming vertical hydrostatic equilibrium, the scale height of the CPD is given by \citep{dempsey2022ContractingExpandingBinary}
\begin{align}
    H_{\rm CPD}\!(r_{\rm cyl})=\frac{H_0}{\sqrt{3}}\Bigg(\frac{r_{\rm cyl}}{R_{\rm H}}\Bigg)^{3/2}\Bigg[1+\frac{1}{3}\Bigg(\frac{r_{\rm cyl}}{R_{\rm H}}\Bigg)^{3}\Bigg]^{-1/2},\label{eq:cpd scale height}
\end{align}
\added[id=R2]{which approximately gives} $\rho_{\rm p}\simeq2\rho_0/(1+0.04\,K)$.
Equation~\ref{eq:mdot locally iso} can be rewritten as
\begin{align}
    \dot{M}_{\rm acc}^{\rm Liso}\!(q)\simeq\frac{2\pi h_0}{1+0.04\,q^2/(\alpha\,h_0^5)}\Bigg(\frac{q}{3}\Bigg)^{2/3}\,\rho_0r_0^3\Omega_0.
\end{align}

\added[id=R1]{
To extend the locally isothermal accretion rate to finite cooling times,
we adopted a critical cooling time derived by \citet{miranda2020PlanetDiskInteraction}
from linear wave theory of planet-driven density waves,}
\begin{align}
    \beta_{\rm crit}=\frac{4}{\gamma-1}h_0^3\simeq1.5\times10^{-3}\,\Bigg(\frac{h_0}{0.05}\Bigg)^3\Bigg|_{\gamma=1.4},\label{eq:beta crit}
\end{align}
below which a locally isothermal condition can be reproduced. 
Combining these relations, we find that the gas accretion rates obtained in our simulations are empirically fitted, \added[id=R4]{with the coefficients determined by a least-squares fit to the simulation data, as
\begin{align}
    \dot{M}_{\rm acc}(q,\beta) =
    \dot{M}_{\rm acc}^{\rm Liso}(q)
    \times
    \min\!\left[
        1,\,
        \left(\frac{\beta}{\beta_{\rm crit}}\right)^{-0.176}
    \right].
    \label{eq:mdot empirical}
\end{align}
Equation~\ref{eq:mdot empirical}} is shown as dashed curves in Fig.~\ref{fig:1d_mdot_vs_qpl}.
The limitations of Eq.~\ref{eq:mdot empirical} are discussed in Sect.~\ref{sec:Model limitations and extensions}.

\added[id=R1]{
We note that Eq.~\ref{eq:mdot empirical} is expressed in code units. 
Adopting a disk model with surface density 
$\Sigma = 10^3\,\mathrm{g\,cm^{-2}}(r/1\,\mathrm{au})^{-1}$ 
and temperature 
$T = 150\,\mathrm{K}\,(r/1\,\mathrm{au})^{-0.5}$, 
together with a fixed opacity of $\kappa=0.1\,\mathrm{cm^2\,g^{-1}}$, 
we obtain 
\added[id=R2]{$\dot{M}_{\rm acc}\simeq1.9\times10^{-5}\,M_{\rm Jup}\,\mathrm{yr^{-1}}$} 
for $\beta\approx10^2$ at 1 au, 
and 
we obtain 
\added[id=R2]{$\dot{M}_{\rm acc}\simeq5.9\times10^{-6}\,M_{\rm Jup}\,\mathrm{yr^{-1}}$}
for $\beta\approx10^0$ at 20 au ($M_{\rm Jup}$ is the Jupiter mass;  Fig.~\ref{fig:cooling_time_contour}).
}

\subsection{Comparison to previous works}\label{sec:Comparison to previous works}
A key feature of our study is the use of long-term, 3D global simulations spanning a wide range of cooling times.
As shown in Sect.~\ref{sec:Numerical results}, inefficient cooling reduces gas accretion by weakening shocks and producing a less dense, more vertically extended CPD.
These effects have been widely discussed in previous work.

\citet{tanigawa2002GasAccretionFlows} investigated accretion flows onto planets using 2D local isothermal simulations.
By varying the normalized sound speed, they showed that both spiral and bow shocks weaken in hotter disks.
This trend has also been confirmed in 2D and 3D simulations including radiative cooling \citep{dangelo2003ThermohydrodynamicsCircumstellarDisks, ayliffe2009GasAccretionPlanetary, szulagyi2016CircumplanetaryDiscCircumplanetary}.
\citet{tanigawa2002GasAccretionFlows} further found that the accretion band narrows and shifts outward as the Mach number decreases.
While the narrowing is consistent with our results (Fig.~\ref{fig:accretion_cross_section}), we do not observe a systematic outward shift, likely reflecting the role of global radial flows captured only in global simulations.

Inefficient cooling also favors the formation of envelope-like structures in the circumplanetary region \citep[e.g.,][]{ayliffe2009CircumplanetaryDiscProperties, fung2019CircumplanetaryDiskDynamics, sagynbayeva2025CircumplanetaryDisksAre}.
\citet{krapp2024ThermodynamicCriterionFormation} derived a thermodynamic criterion for CPD formation from 3D global multifluid (gas+dust) radiation hydrodynamic simulations,
$t_{\rm cool}/(2\pi/\Omega_0)\lesssim0.1$ ($\beta\lesssim0.6$), in qualitative agreement with our results (Fig.~\ref{fig:2d_slice_local_sigma_rho}).

Despite extensive work on thermodynamic effects, only a limited number of simulations have followed gas accretion onto gap-opening planets for sufficiently long times to reach a quasi-steady state while accounting for radiative cooling.
Most studies are restricted to 2D setups, short integration times (tens to hundreds of orbits), or a narrow range of cooling times.

Nevertheless, even in isothermal simulations, varying the sound speed provides useful insight into how the characteristic Mach number of the flow regulates gas accretion.
In 2D local simulations of \citet{tanigawa2002GasAccretionFlows}, the authors obtained an empirical relation between the gas accretion rate and the sound speed, $\dot{M}_{\rm acc}\propto c_{\rm s}^{-2}$, consistent with our finding that hotter (lower Mach number) conditions lead to reduced gas accretion rates.
Below, we further examine the extent to which this picture is supported by hydrodynamical simulations with radiative cooling.

\citet{ayliffe2009GasAccretionPlanetary} performed 3D global radiation hydrodynamic simulations and found that Jupiter-mass planets accrete at rates similar to locally isothermal predictions, largely independent of opacity.
However, their accretion rates were measured at 10 orbits, likely before reaching a quasi-steady state, suggesting that the suppression of gas accretion may not manifest on short timescales.
Recent 3D global radiation hydrodynamic simulations extending over several tens of orbits identified a clear opacity dependence of the accretion rate \citep{schulik2019Global3DRadiationhydrodynamic, schulik2020StructureMassDelivery, lambrechts2019QuasistaticContractionRunaway}, with \citet{schulik2019Global3DRadiationhydrodynamic} empirically finding $\dot{M}_{\rm acc}\propto\kappa^{-1/4}$.

Long-term simulations tracking the gas accretion process while accounting for cooling have primarily been conducted using the $\beta$-cooling approximation.
\added[id=R1]{Using a local setup with prescribed gap profile,} \citet{zhu2016ShockdrivenAccretionCircumplanetary} carried out 3D simulations with $\beta$ cooling and showed that inefficient cooling produces a less dense and more vertically extended CPD, together with a reduced gas accretion rate, in agreement with our results.
\citet{wu2024EffectsThermodynamicsConcurrent} performed 2D global simulations with $\beta$ cooling for a planet with $q=10^{-3}$, systematically varying $\beta$ over the range $\beta\in[10^{-2},10^{2}]$, directly comparable to our study. 
They reported only a modest decrease in the accretion rate with increasing $\beta$, within a factor of two (see their Fig.~1), and a similarly weak reduction in the CPD surface density within $r_{\rm cyl}<0.1\,R_{\rm H}$.
In contrast, our 3D simulations show that the gas accretion rate can be reduced by up to an order of magnitude at large $\beta$.
This difference highlights the intrinsic importance of 3D effects for quantifying gas accretion, as 2D simulations cannot capture either the narrowing of the accretion band at high altitude or the increase in the CPD scale height.
\added[id=R4]{Part of the discrepancy may also arise from differences in the adopted planetary potential. 
The effective potential in vertically integrated 2D disks is not generally equivalent to a Plummer potential, but is closer to a Bessel potential \citep{muller2012TreatingGravityThindisk,brown2024HorseshoesSpiralWaves,cordwell2025HowTwodimensionalAre}. 
How the choice of effective potential influences gas accretion rates remains unclear and should be explored in future work.}

Finally, we assess the dependence of the gas accretion rate on the surface density in the gap region \citep{tanigawa2007SystematicStudyFinal, tanigawa2016FinalMassesGiant, lega2024GasDynamicsJupitermass}.
Recent studies have shown that both the width and depth of gaps vary with the cooling time, $\beta$ \citep{miranda2020PlanetDiskInteraction, zhang2024DependenceStructurePlanetopened, ono2025ModelingProtoplanetaryDisk}, consistent with our results (Fig.~\ref{fig:1d_plot_sigma}a).
However, we find that these $\beta$-dependent variations in the gap profile have little impact on the gas accretion rate.
Instead, the accretion rate is primarily regulated by the CPD scale height and by the cross section of the accretion band.
The CPD scale height is largely independent of the gap profile.
Although the accretion band is located within the gap, at \added[id=R4]{$|r|\simeq r_0+2\,R_{\rm H}$}, and could therefore be influenced by gap properties, the surface density at its location varies only weakly with $\beta$ (Fig.~\ref{fig:1d_plot_sigma}b).
This implies that the CPD surface density is controlled mainly by the $\beta$-dependent shock structure, rather than by $\beta$-dependent variations in the gap profile.

\subsection{Implications for observations}

For a fixed opacity, the ambient cooling time increases toward the inner disk (Fig.~\ref{fig:cooling_time_contour}), rendering gas accretion onto planets inefficient in the inner, hotter regions and prolonging mass-doubling timescales.
This trend may be consistent with the relatively low occurrence rate of hot Jupiters inferred from the observed mass–period distribution of exoplanets \citep{johnson2010GiantPlanetOccurrence, fernandes2019HintsTurnoverSnow, fulton2021CaliforniaLegacySurvey, lagrange2023RadialDistributionGiant}.
Further studies are nevertheless required to assess the impact of cooling-regulated accretion on planetary growth tracks while accounting for additional physics, such as disk evolution, planet migration, and spatial variations in the cooling time.

PDS~70 is a well-studied system hosting accreting planets, for which the masses of PDS~70b and c have been inferred from observations \citep{keppler2018DiscoveryPlanetarymassCompanion, muller2018OrbitalAtmosphericCharacterization, benisty2021CircumplanetaryDiskPDS70c, wang2021ArchitecturePlanetarySystems, portilla-revelo2023ConstrainingGasDistribution}.
Gas accretion rates, or the product $M_{\rm p}\dot{M}_{\rm acc}$, have also been extensively estimated \citep{wagner2018MagellanAdaptiveOpticsa,haffert2019TwoAccretingProtoplanetsa,
thanathibodee2019MagnetosphericAccretionSource,aoyama2019ConstrainingPlanetaryGas, christiaens2019EvidenceCircumplanetaryDisk,stolker2020MIRACLESAtmosphericCharacterization, zhou2021HubbleSpaceTelescope, shibaike2024ConstraintsPDS70}.
\added[id=R2]{
Recently, the WIde Separation Planets In Time (WISPIT) survey has detected a young accreting protoplanet, WISPIT~2b, via H$\alpha$ emission \citep{vancapelleveen2025WIdeSeparationPlanets}.  
Among these planets, PDS~70c and WISPIT~2b exhibit accretion rates that are expected to be higher by a factor of approximately $2$ compared to PDS~70b \citep{close2025WideSeparationPlanets}.  
These planets are located at larger orbital separations \citep[PDS~70b: 20 au, PDS~70c: 34 au, WISPIT~2b: 55 au;][]{haffert2019TwoAccretingProtoplanetsa,vancapelleveen2025WIdeSeparationPlanets}, where the disk conditions may correspond to lower $\beta$ (Fig.~\ref{fig:cooling_time_contour}), potentially leading to higher $\dot{M}_{\rm acc}$, in broad agreement with our results.
}

\added[id=R3]{More recently, WISPIT~2c has been identified, with no H$\alpha$ emission detected \citep{lawlor2026DirectSpectroscopicConfirmation}.  
This absence of emission may indicate that a large-scale envelope surrounding the planet obscures the accretion signal. 
At its orbital distance of 14 au, where disk conditions may correspond to higher $\beta$, the presence of such an envelope is consistent with our simulations (Fig.~\ref{fig:2d_slice_local_sigma_rho}f).}

When either the planet mass or the gas accretion rate is independently estimated, albeit with uncertainties, a theoretical $M_{\rm p}$--$\dot{M}_{\rm acc}$ relation---often derived under the locally isothermal assumption---can be used to assess their mutual consistency.
Our results emphasize that the additional degree of freedom introduced by the cooling time can lead to systematic uncertainties, rendering the interpretation of observational data sensitive to disk thermal conditions.
\added[id=R4]{Beyond the measurements of $M_{\rm p}$ or $\dot{M}_{\rm acc}$, cooling-time-dependent changes in the circumplanetary structure may also modify the broader observational appearance of embedded planets, including their spectral energy distributions
\citep{taylor2023AwesomeSOSSAtmospheric, choksi2025SpectralEnergyDistributions}.}
Efforts to constrain cooling times through opacity measurements \citep{testi2014DustEvolutionProtoplanetary, birnstiel2018DiskSubstructuresHigh} or disk-kinematic estimates \citep{hall2020PredictingKinematicEvidence, longarini2021InvestigatingProtoplanetaryDisk} will therefore be crucial.

\added[id=R1]{
Velocity perturbations in disks can serve as imprints of forming planets and have been revealed by recent observations \added[id=R4]{\citep[][]{pinte2018KinematicEvidenceEmbedded, dong2019ObservationalSignaturesPlanets, izquierdo2023DiscMinerIIRevealing, izquierdo2026ExoALMAXXTomographic}.}
In particular, vertical flows toward the midplane with speeds of approximately 10\% of the sound speed at a few scale heights above the midplane have been detected in the disk around HD~163296, and interpreted as signatures of large-scale gas flows driven by Jupiter-mass planets \citep{teague2019MeridionalFlowsDisk}.}

\added[id=R1]{
Our results indicate that the velocity structure, especially its vertical component, is cooling-time dependent.
As shown in Sect.~\ref{sec:Cooling-regulated mass supply to the CPD}, the horseshoe streamlines of non-accreting gas do not significantly change their altitude under nearly isothermal conditions ($\beta=10^{-2}$; Fig.~\ref{fig:3d_streamlines_b1e-02}), whereas a prominent vertical motion develops in the trailing horseshoe region under nearly adiabatic conditions ($\beta=10^{2}$; Fig.~\ref{fig:3d_streamlines_b1e+02}).
Figure~\ref{fig:2d_slice_vz} further shows that the vertical velocity at $z=2H$ in the corotation region reaches approximately 20--30\% of the sound speed.
This vertical motion likely arises from baroclinic vorticity generation associated with the development of a superadiabatic temperature gradient \citep[cf. Fig.~5 of][]{chrenko2019OscillatoryMigrationAccretinga}.
We note that the vertical motion becomes apparent only after the gap is well developed (after approximately $300$ orbits), which may help reconcile our results with shorter-duration simulations \citep{fung2019CircumplanetaryDiskDynamics}, who reported no significant vertical variations in the horseshoe streamlines in their adiabatic runs.
Future work should explore this cooling-dependent velocity structure and compare it quantitatively with molecular line observations.
}

\added[id=R4]{
The thermodynamic structure revealed in our simulations may also have implications for the chemical signatures of forming planets in disks.
The temperature enhancement around the planet becomes more pronounced as $\beta$ increases (Fig.~\ref{fig:2d_slice_temperature}).
Volatile species may sublimate within these planet-induced high-temperature regions  \citep{cleeves2015INDIRECTDETECTIONFORMING, jiang2023ChemicalFootprintsGiant}, which may help interpret molecular emission features observed near embedded planets \citep[e.g.,][]{law2021MoleculesALMAPlanetforminga, booth2023SulphurMonoxideEmission}.
Since the extent of the high-temperature region depend sensitively on the cooling time, the conditions under which such chemical footprints emerge may likewise depend on the cooling efficiency of the disk.
}
\subsection{Model limitations and extensions}\label{sec:Model limitations and extensions}
\paragraph{$\beta$ cooling approximation.}
The $\beta$-cooling model employed in this study is a simplified treatment of radiative cooling and therefore has several limitations \citep[e.g.,][]{gammie2001NonlinearOutcomeGravitational}.
In disk–planet interaction simulations, numerical instabilities can arise at the inner boundary for extremely large $\beta$ values \citep[$\beta>10^2$;][]{ono2025ModelingProtoplanetaryDisk, okuzumi2026BridgingGapConsistent}.
The $\beta$ values adopted here remain below this threshold, and we therefore do not observe the instability.
Such large $\beta$ values may nevertheless be realized in the inner disk (Fig.~\ref{fig:cooling_time_contour}), warranting caution when applying Eq.~\ref{eq:mdot empirical} at very large $\beta$.

$\beta$-cooling simulations can overestimate the gap-opening efficiency relative to full radiative transfer, particularly for intermediate cooling times \citep[$\beta\sim1$;][]{ziampras2026HowTwodimensionalAre}.
As discussed in Sect.~\ref{sec:Comparison to previous works}, our results indicate that the gas accretion rate is regulated primarily by the $\beta$-dependent shock structure and the CPD scale height, rather than by the detailed gap profile.
From this perspective, we expect the gas accretion rates obtained in this study to remain qualitatively robust.
Nevertheless, it remains uncertain whether the shock strengths and thermodynamic structures produced in $\beta$-cooling simulations quantitatively reproduce those obtained in radiation hydrodynamic models.
Direct comparisons with radiation hydrodynamic simulations will therefore be required to fully assess the robustness of cooling-regulated gas accretion.

Despite these limitations, the $\beta$-cooling model remains a practical tool for broad parameter surveys owing to its substantially lower computational cost compared to full radiation hydrodynamic simulations.
For simplicity, we adopted a fixed $\beta$ throughout the computational domain, although $\beta$ is expected to vary spatially in realistic disks \added[id=R4]{\citep{zhu2015StructureSpiralShocks, baehr2015RoleCoolingPrescription,takahashi2016RevisedConditionSelfgravitational,miranda2020PlanetDiskInteraction, bae2021ObservationalSignatureTightly, ziampras2023ModellingPlanetinducedGaps}.}
In particular, the local cooling time in the Hill and gap regions likely plays a key role in setting the CPD scale height and the width of the accretion band \citep{zhu2015StructureSpiralShocks, krapp2024ThermodynamicCriterionFormation}.
\added[id=R4]{An adaptive $\beta$-cooling prescription could also introduce an asymmetry in the accretion flow, since the cooling time may vary across the planetary orbit at the location of the accretion bands (Fig.~\ref{fig:cooling_time_contour}).}
Because dust opacity is a primary source of the local cooling time, these considerations motivate future high-resolution studies that self-consistently couple dust evolution and radiative cooling.

\paragraph{The sink cell approach.}
\added[id=R1]{
To measure the gas accretion rate onto the planet, we adopted a sink cell approach that has been widely used in previous studies \citep[e.g.,][]{bryden2000InteractionProtoplanetsProtostellarDisks, nelson2000MigrationGrowthProtoplanetsa, choksi2023MaximumAccretionRate}. 
Gas entering a sphere of radius $0.1\,R_{\rm H}$ was removed on a prescribed timescale.
Because this sink radius greatly exceeds the physical planetary radius, the measured accretion rate corresponds to the mass flux entering $<0.1\,R_{\rm H}$ rather than to thermally regulated contraction onto the planetary surface \citep{lambrechts2019QuasistaticContractionRunaway}.
We also adopted a gravitational smoothing length of $0.1\,R_{\rm H}$ \added[id=R2]{(and $0.05\,R_{\rm H}$ for a convergence test; see Appendix~\ref{sec:Convergence tests})}, such that the immediate vicinity of the planet remains unresolved in our global simulations.
In particular, luminosity-regulated accretion is not treated self-consistently, as discussed below.
Nevertheless, local high-resolution simulations resolving the planetary radius have shown that, in the runaway regime, most gas entering $\lesssim 0.5\,R_{\rm H}$ ultimately accretes onto the planetary surface \citep{machida2010GasAccretionProtoplanet}. 
We therefore interpret the measured accretion rates as supply-limited rates appropriate to the runaway phase explored here.}
\added[id=R4]{We note, however, that the measured accretion rate depends only weakly on the sink prescription in the fast-cooling regime, while a larger systematic uncertainty remains in the slow-cooling regime (Appendix~\ref{sec:Convergence tests}).}

\paragraph{Effects of planet luminosity.}
The accretion of solids and gas can generate intrinsic luminosity \citep{lambrechts2017ReducedGasAccretion, ginzburg2019EndgameGasGianta}, which is not included in our setup.
Near the sink radius, such accretion heating may lead to self-regulation of gas accretion by inflating the CPD \citep{tanigawa2002GasAccretionFlows}, potentially further suppressing gas accretion.

\paragraph{Dependence on planet mass.}
This study focused on planets with $q=10^{-3}\text{--}3\times10^{-3}$, corresponding to approximately 1 to 3 Jupiter masses around a solar-mass star.
It is nontrivial whether the empirical formula (Eq.~\ref{eq:mdot empirical}) can be extended to lower planet masses.
In the low-mass regime, disk gas approaches the planet subsonically owing to the narrowing of the corotation region \citep{ormel2015HydrodynamicsEmbeddedPlanets, fung20153DFlowField, kuwahara2019GasFlowPlanet}, suppressing shock dissipation and potentially leading to a qualitatively different $\beta$-dependence of the accretion rate.

\paragraph{Implications for planet migration.}
Planet migration is a long-standing problem in planet formation theory.
Recent studies have shown that gas accretion can qualitatively alter the direction of migration.
In disks with $\alpha \gtrsim 3\times10^{-3}$, higher than the assumed $\alpha$ value in this work, Jupiter-mass planets undergoing gas accretion tend to migrate outward, whereas non-accreting planets migrate inward.
This behavior has been attributed to asymmetric spiral structures connected to the CPD \added[id=R1]{in locally isothermal simulations} \added[id=R4]{\citep{li2024ConcurrentAccretionMigration, pan2026ConcurrentAccretionMigration, ida2026OutwardMigrationGas}}.
In contrast, in slow-cooling regimes ($\beta \gtrsim 1$), thermodynamic effects substantially modify the CPD structure, and the tendency toward inward migration may be recovered, as suggested by 2D global simulations employing a $\beta$-cooling prescription \citep{wu2024EffectsThermodynamicsConcurrent}.

\added[id=R4]{Migration may also differ in lower-viscosity disks.
A gap-opening planet is expected to undergo type-II migration, but at sufficiently low viscosities, $\alpha\lesssim10^{-4}$, vortices formed at the gap edges can exert time-dependent torques on the planet, leading to transient migration jumps or stalling \citep{mcnally2019MigratingSuperEarthsLowviscosity, lega2021MigrationJupitermassPlanets, lega2022MigrationJupiterMass, weiskopf2026FormationMultipleDust}.
Appendix~\ref{sec:Torque analysis} finds that, for $\alpha=10^{-3}$, the total torque remains negative for all cooling times explored here, implying an initial tendency toward inward migration.
Future work should explore the combined effects of radiative cooling and turbulence on planet migration in 3D.}

\paragraph{Midplane asymmetry}
\added[id=R4]{Our simulations were restricted to the upper half of the disk by imposing midplane symmetry. 
For the physical setup considered in this work, we do not expect this approximation to substantially affect the gas accretion rate onto the planet. 
However, additional physics not included here, such as magnetic fields, inclined planetary orbits, asymmetric irradiation, or shadowing effects, could break the midplane symmetry and lead to asymmetric accretion flows \citep{nealon2018WarpingProtoplanetaryDisc,nealon2019ScatteredLightShadows,hu20253DGapOpening,zhang2025ShadowinducedWarpsProtoplanetary}. 
Exploring such asymmetric configurations requires dedicated full 3D simulations.
}

\section{Conclusions}\label{sec:Conclusions}
We have conducted a suite of three-dimensional, long-term hydrodynamical simulations to quantify gas accretion rates onto gap-opening planets over a wide range of cooling conditions.
By varying the dimensionless cooling timescale, $\beta$, we find that the gas accretion rate, $\dot{M}_{\rm acc}$, systematically decreases with increasing $\beta$.
When the cooling time is 100 times the Keplerian time ($\beta=10^2$), the quasi-steady accretion rate is reduced by approximately an order of magnitude relative to locally isothermal predictions.

The suppression of gas accretion at large $\beta$ originates from thermodynamic effects in the circumplanetary region.
Inefficient cooling weakens bow and spiral shocks, reducing energy dissipation at the shocks and thereby narrowing the accretion bands that supply gas to the CPD. \added[id=R1]{Additionally,} higher temperatures inflate the CPD vertically.
Together, these effects lower the gas volume density in the CPD and reduce the gas accretion rate onto the planet.
We further find an empirical scaling relation, \added[id=R4]{$\dot{M}_{\rm acc}\propto\beta^{-0.176}$}, which extends locally isothermal accretion models by explicitly accounting for cooling effects.

Finally, our findings highlight the importance of thermodynamic effects when interpreting observations of accreting planets.
Theoretical $M_{\rm p}$--$\dot{M}_{\rm acc}$ relations, often derived under locally isothermal assumptions, may suffer from systematic uncertainties if cooling times are not properly accounted for.
Moreover, the cooling-regulated accretion bands identified here suggest a possible link between gas accretion and the delivery of solids to CPDs, motivating future coupled gas--dust studies.

\begin{acknowledgements}
\added[id=R4]{We would like to thank the anonymous referee for the constructive report, which helped us improve the clarity and robustness of the manuscript.}
We thank Athena++ developers.
Numerical computations were carried out on HPE Cray XD2000 at the Center for Computational Astrophysics, National Astronomical Observatory of Japan \added[id=R4]{and the Tycho supercomputer hosted at the SCIENCE HPC center at the University of Copenhagen}.
\added[id=R2]{We thank Takayuki Tanigawa, Yuhito Shibaike, and Alexandros Ziampras for helpful discussions, and Ya-Ping Li for useful comments on the sink cell implementation.}
M.L. acknowledge the ERC starting grant 101041466-EXODOSS.
\end{acknowledgements}
%


\bibliography{zotero_references}

\clearpage
\begin{appendix}

\def\thesection{A}
\setcounter{equation}{0}
\def\theequation{A.\arabic{equation}}
\setcounter{figure}{0}
\def\thefigure{A.\arabic{figure}}

\section{Convergence tests}\label{sec:Convergence tests}
\begin{figure}[tp]
    \centering
    \includegraphics[width=1\linewidth]{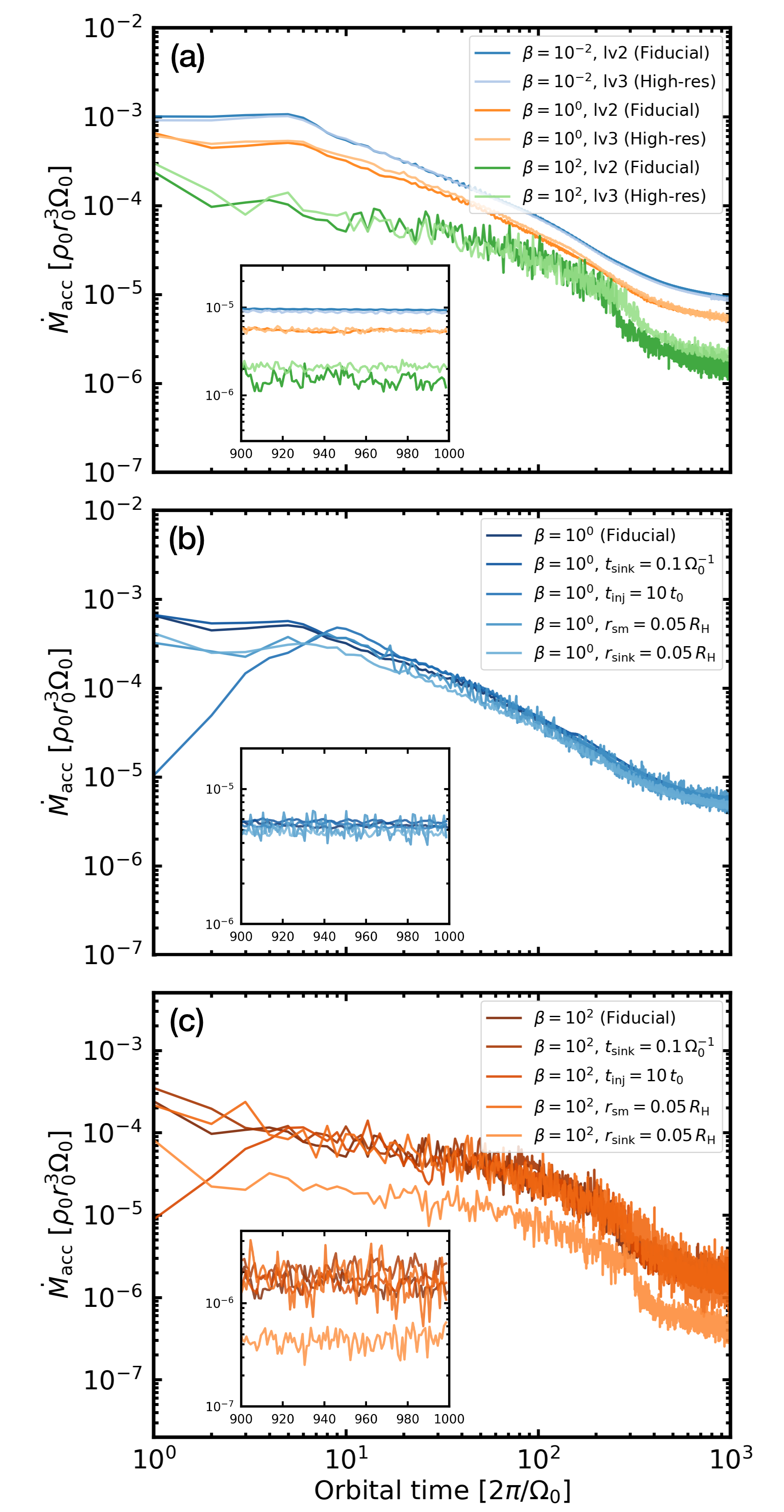}
    \caption{Time evolution of the gas accretion rate onto a planet with $q=10^{-3}$ for different numerical settings \added[id=R2]{(Table~\ref{tab:hydro simulations}). The inset shows a zoom-in of the last 100 orbits on linear scale. \textit{Top:} dependence of the accretion rate on the resolution within the Hill region. \added[id=R4]{\textit{Middle and bottom:} dependence on the sink timescale, the injection timescale, the smoothing length, and the sink radius.}}}
    \label{fig:1d_plot_mdot_convergence_test}
\end{figure}

\begin{figure}[tp]
    \centering
    \includegraphics[width=1\linewidth]{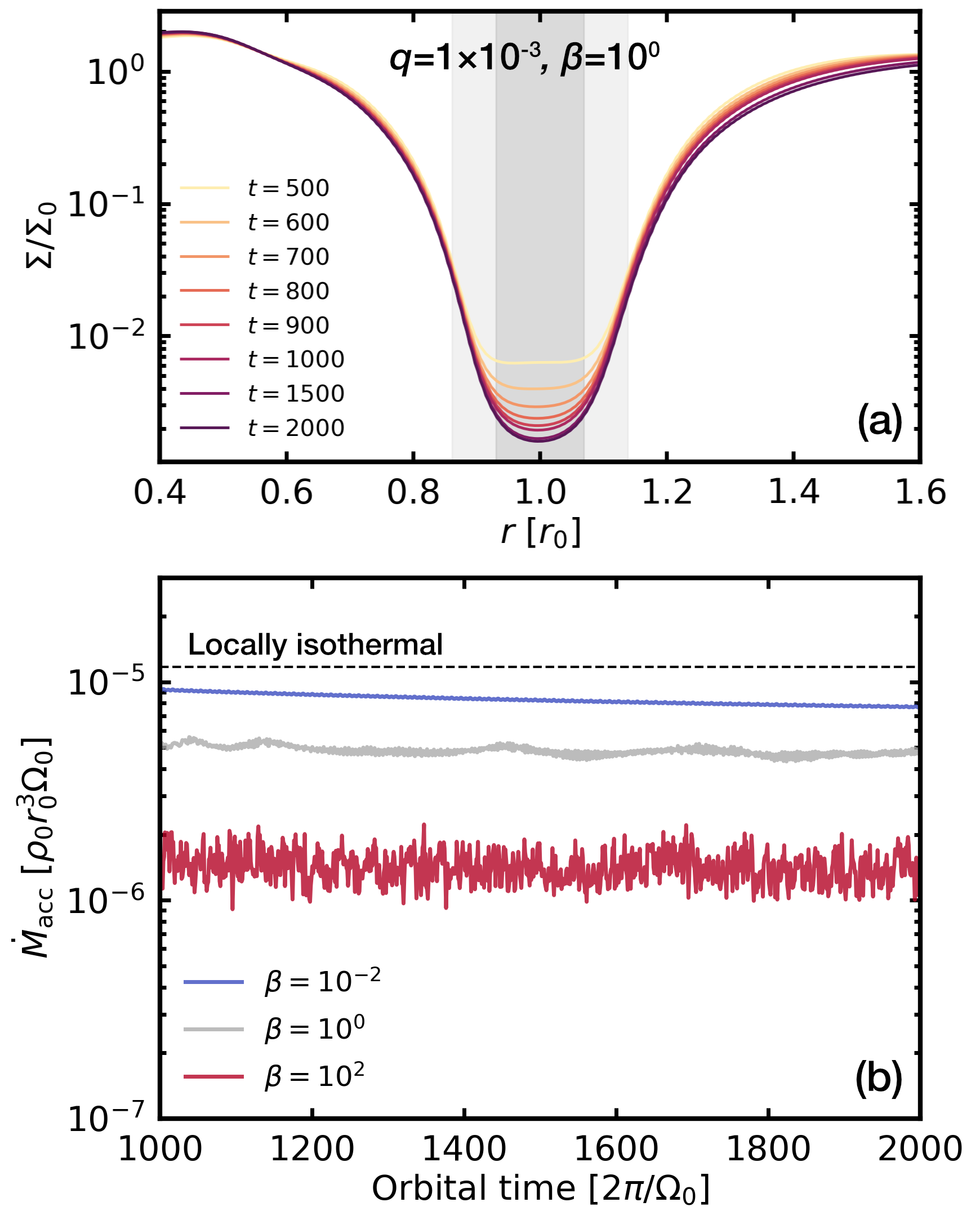}
    \caption{\added[id=R4]{Time evolution of the gas surface density (\textit{top}) and the long-term evolution of the gas accretion rate (\textit{bottom}). Dark and light gray shaded regions in the top panel indicate $r=r_0\pm R_{\rm H}$ and $r=r_0\pm 2\,R_{\rm H}$, respectively. The bottom panel shows the evolution after $10^3$ orbits. The horizontal dashed line shows the prediction of a locally isothermal model (Eq.~\ref{eq:mdot locally iso}).}}
    \label{fig:longer_run}
\end{figure}

This section explores how the simulation results depend on numerical settings: the resolution within the Hill region, the sink radius, \added[id=R2]{the sink timescale, the smoothing length, the injection timescale, \added[id=R4]{and the integration time}}. 
Figure~\ref{fig:1d_plot_mdot_convergence_test} shows that the quasi-steady gas accretion rates converge to similar values across different numerical settings.

\added[id=R4]{The only notable exception is the run with the smaller sink radius for $\beta=10^2$, where the accretion rate is lower by approximately a factor of two.
This likely reflects the pressure-supported envelope formed in the slow-cooling regime, which makes the accretion rate more sensitive to how the sink region is defined.
Because the appropriate $r_{\rm sink}$ is not unique \citep[e.g.,][]{tanigawa2002GasAccretionFlows, machida2010GasAccretionProtoplanet}, we regard this difference as a systematic uncertainty associated with the sink prescription.}

\added[id=R4]{
Figure~\ref{fig:longer_run} shows the time evolution of the gas surface density and gas accretion rate beyond $10^3$ orbits, which is the fiducial integration time adopted in this study.
The gap profile reaches a quasi-steady state after $\gtrsim10^3$ orbits, consistent with the analytical estimate of the gap-opening timescale (Eq.~\ref{eq:t_gap}).
The gas accretion rate continues to decrease only slightly after $10^3$ orbits, compared to the much stronger evolution before $10^3$ orbits.
The values measured at $2\times10^3$ orbits are within $10\%$ of those measured at $10^3$ orbits.
Longer integration times also enter the regime, not considered here, where the accreted gas mass into the sink should start to be taken into account.
}

\def\thesection{B}
\setcounter{equation}{0}
\def\theequation{B.\arabic{equation}}
\setcounter{figure}{0}
\def\thefigure{B.\arabic{figure}}

\section{Cooling time estimation}\label{sec:Cooling time estimation}

The cooling time of the background gas is determined by two components \citep{miranda2020PlanetDiskInteraction, ziampras2023ModellingPlanetinducedGaps}: surface cooling (radiative cooling from the disk surface) and in-plane cooling (radial thermal diffusion through the disk midplane). This is given by \citep{ziampras2023ModellingPlanetinducedGaps}:
\begin{align}
    &\beta_{\rm surf}\equiv\frac{\Sigma_0\,c_{\rm V}}{2\sigma_{\rm SB}T_0^3}\tau_{\rm eff}\,\Omega,\quad\tau_{\rm eff}=\frac{3\tau}{8}+\frac{\sqrt{3}}{4}+\frac{1}{4\tau},\quad\tau=\frac{\kappa\Sigma_0}{2},\\
    &\beta_{\rm mid}\equiv\frac{3\kappa\rho_0c_{\rm V}}{16\sigma_{\rm SB}T_0^3}\Bigg(H^2+\frac{l_{\rm rad}^2}{3}\Bigg)\,\Omega,\quad l_{\rm rad}=\frac{1}{\kappa\rho_0},\\
    &\beta_{\rm disk}\equiv\frac{\beta_{\rm surf}}{1+\beta_{\rm surf}/\beta_{\rm mid}}.
    \label{eq:beta disk}
\end{align}
Here $\Sigma_0$ is the gas surface density, $T_0$ is the midplane temperature, $\rho_0=\Sigma_0/(\sqrt{2\pi}H)$ is the gas density, $c_{\rm V}=k_{\rm B}/(\mu m_{\rm H}(\gamma-1))$ is the specific heat capacity, $\kappa$ is the opacity, and $\sigma_{\rm SB}$ is the Stefan-Boltzmann constant. Equation \ref{eq:beta disk} accounts for both optically thick and optically thin regimes.

\def\thesection{C}
\setcounter{equation}{0}
\def\theequation{C.\arabic{equation}}
\setcounter{figure}{0}
\def\thefigure{C.\arabic{figure}}

\begin{figure*}[htp]
    \centering
    \includegraphics[width=1\linewidth]{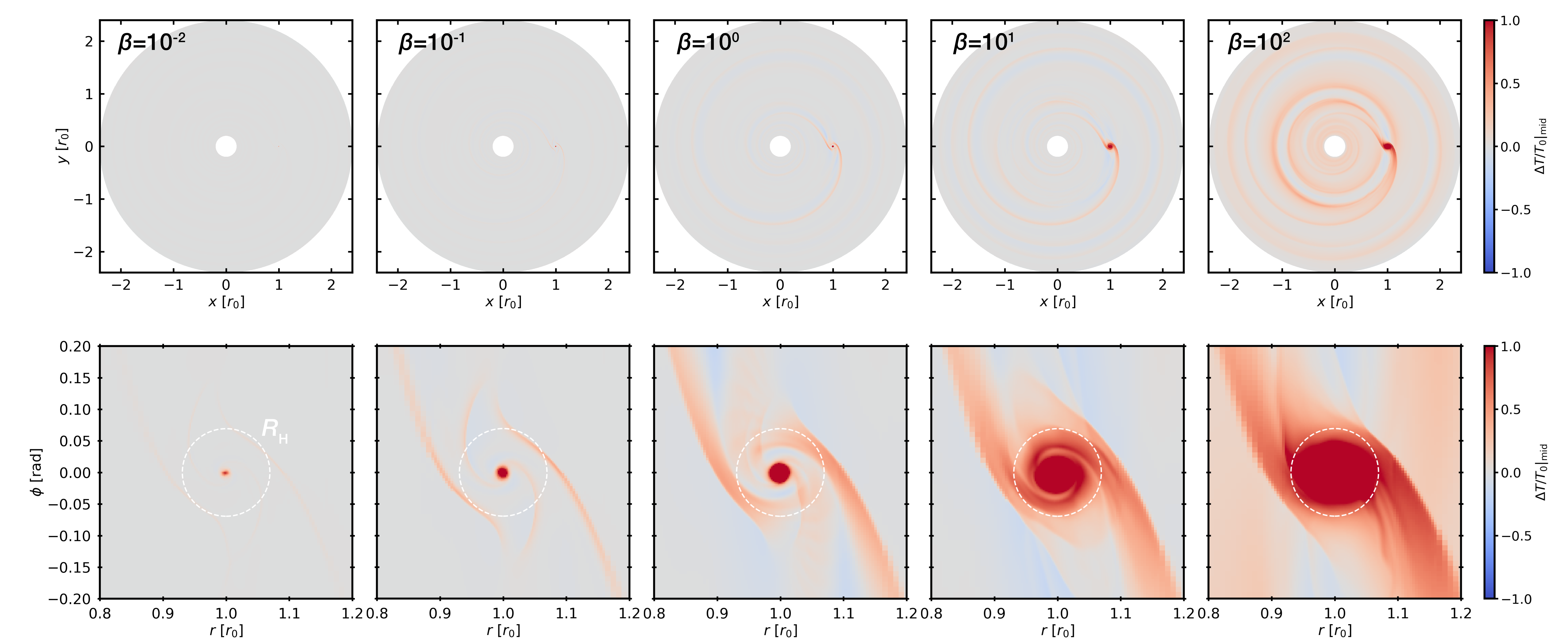}
    \caption{Temperature perturbations at the disk midplane, $\Delta T/T_0|_{\rm mid} = (T - T_0)/T_0|_{\rm mid}$, for different cooling times, $\beta$. The planet with $q=10^{-3}$ is located at $(x,y)=(1,0)$. All panels are snapshots at the end of the simulation, $t=10^3\,t_0$. \textit{Top:} Entire view. \textit{Bottom:} Zoomed view of the circumplanetary region. The dashed circle denotes the Hill radius.}
    \label{fig:2d_slice_temperature}
\end{figure*}

\begin{figure*}[tp]
    \centering
    \includegraphics[width=1\linewidth]{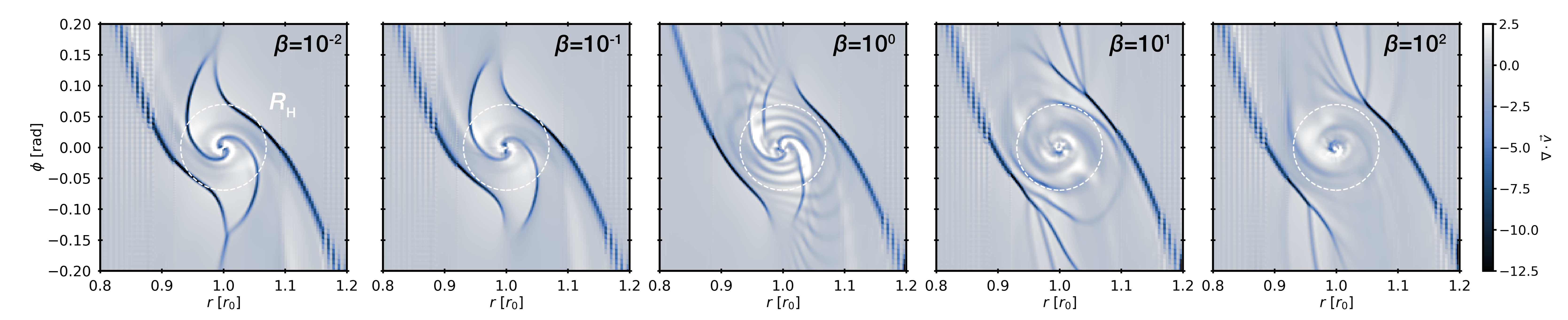}
    \caption{\added[id=R2]{Divergence of the velocity field for different cooling times, $\beta$. Darker colors indicate the locations of shock fronts.}}
    \label{fig:2d_slice_local_div_v}
\end{figure*}

\begin{figure}[tp]
    \centering
    \includegraphics[width=1\linewidth]{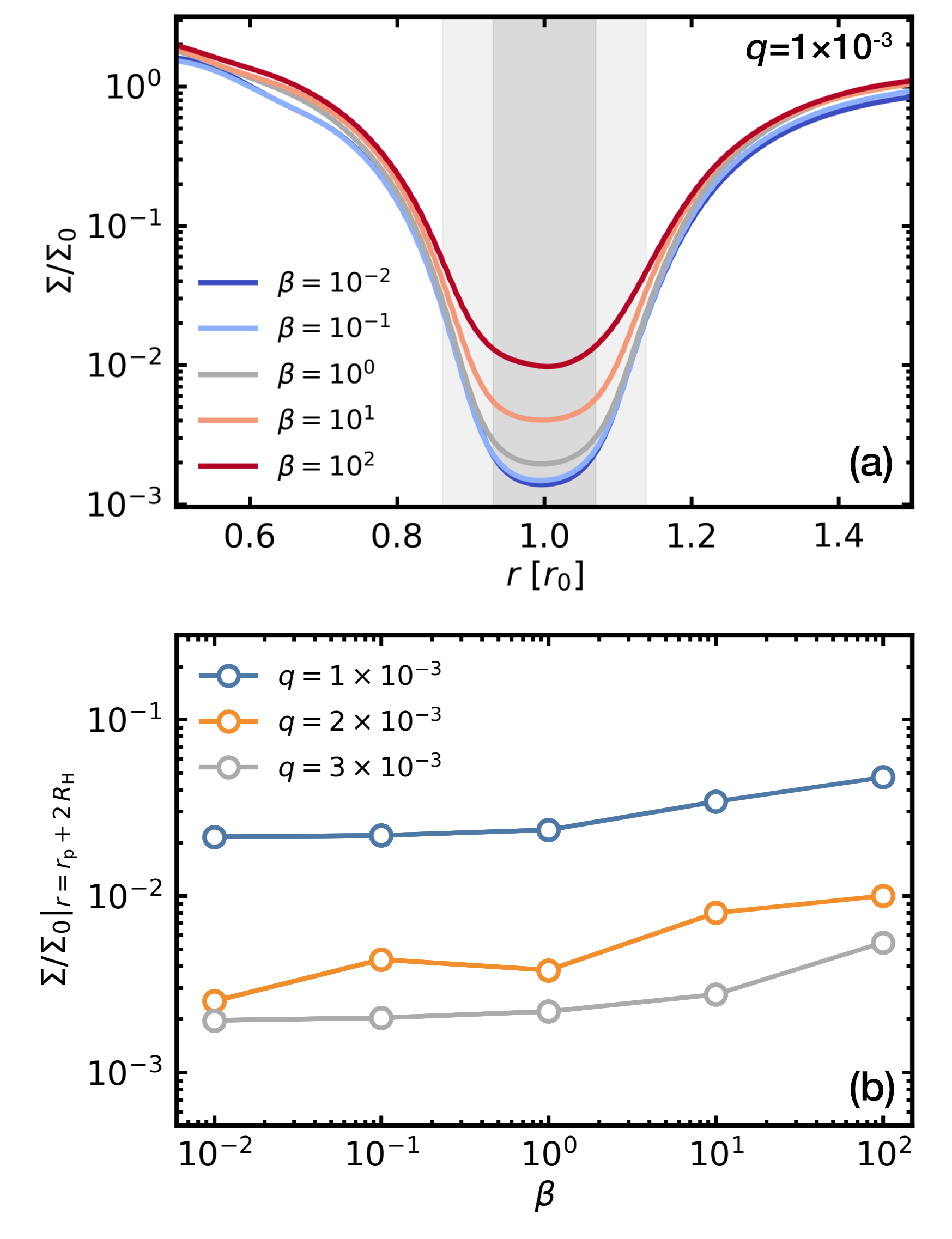}
    \caption{Azimuthally averaged radial profiles of the gas surface density (\textit{top}) and the surface density measured at \added[id=R4]{$r=r_0+2\,R_{\rm H}$} as a function of the dimensionless cooling time $\beta$. \added[id=R4]{All panels are snapshots at $t=10^3\,t_0$.} \textit{Top:} The planet with $q=10^{-3}$ is located at $r=1$. Dark and light gray shaded regions indicate \added[id=R4]{$r=r_0\pm R_{\rm H}$ and $r=r_0\pm 2\,R_{\rm H}$}, respectively. The accretion bands supplying gas to the CPD are located near \added[id=R4]{$r\simeq r_0\pm 2\,R_{\rm H}$} (see Fig.~\ref{fig:accretion_bands}).}
    \label{fig:1d_plot_sigma}
\end{figure}

\begin{figure}[tp]
    \centering
    \includegraphics[width=1\linewidth]{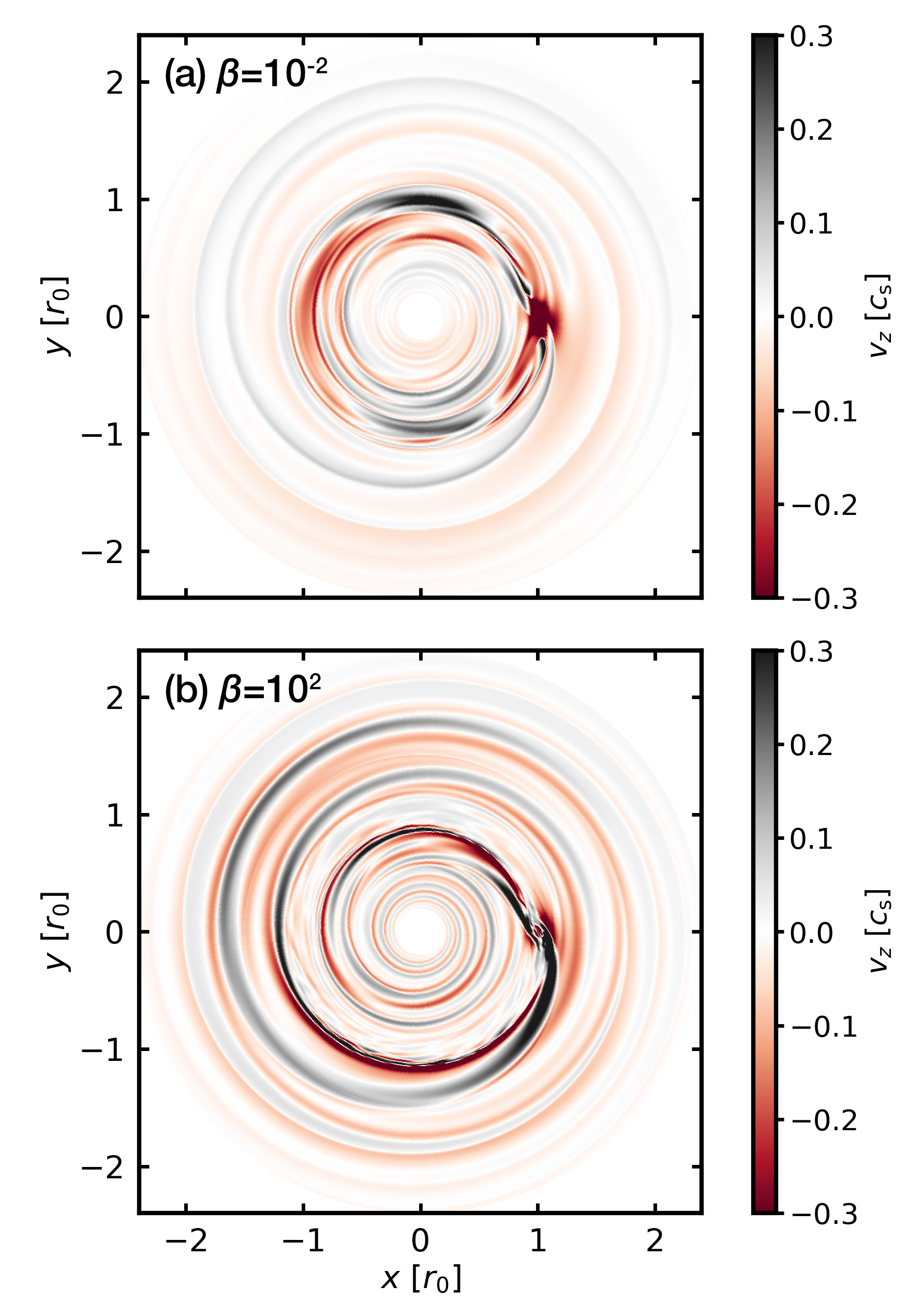}
    \caption{Vertical velocity at $z=2\,H$ normalized by the local sound speed. All panels are snapshots at the end of the simulation, $t=10^3\,t_0$. The planet with $q=10^{-3}$ is located at $(x,y)=(1,0)$.}
    \label{fig:2d_slice_vz}
\end{figure}

\section{\added[id=R4]{Dependence on cooling time}}
Increasing the cooling time leads to higher gas temperatures in the circumplanetary region.
Figure~\ref{fig:2d_slice_temperature} shows the temperature perturbations at the disk midplane.
Temperature perturbations are negligible throughout the disk when $\beta=10^{-2}$, but become prominent near the planet and along the density waves as $\beta$ increases. 
\added[id=R4]{Figure~\ref{fig:2d_slice_local_div_v} shows the divergence of the velocity field of the gas at the midplane.
The darker color indicates the shock locations.}
Figure~\ref{fig:1d_plot_sigma} shows the radial profile of the surface density, as well as the surface density measured at \added[id=R4]{$r = r_0 + 2\,R_{\rm H}$}.
Following \citet{zhang2024DependenceStructurePlanetopened}, the surface density was azimuthally averaged while excluding the azimuthal region around the planet, $\phi \in [-\pi/12,\,\pi/12]$.
\added[id=R1]{Figure~\ref{fig:2d_slice_vz} shows the vertical velocity at $z=2\,H$ for different cooling times, $\beta=10^{-2}$ (panel a) and $\beta=10^{2}$ (panel b).
A clear difference emerges near the planet and in the corotation region.
For $\beta=10^{-2}$, a downward velocity associated with high-altitude accretion flows is visible at the planet’s location.
In contrast, for $\beta=10^{2}$, gas does not accrete efficiently from high altitudes (Fig.~\ref{fig:accretion_bands}), and no prominent downward motion is observed near the planet.
Instead, a strong upward flow develops in the corotation region, reaching approximately 20--30\% of the local sound speed.
}

\def\thesection{D}
\setcounter{equation}{0}
\def\theequation{D.\arabic{equation}}
\setcounter{figure}{0}
\def\thefigure{D.\arabic{figure}}

\section{Torque analysis}\label{sec:Torque analysis}

\begin{figure}[tp]
    \centering
    \includegraphics[width=1\linewidth]{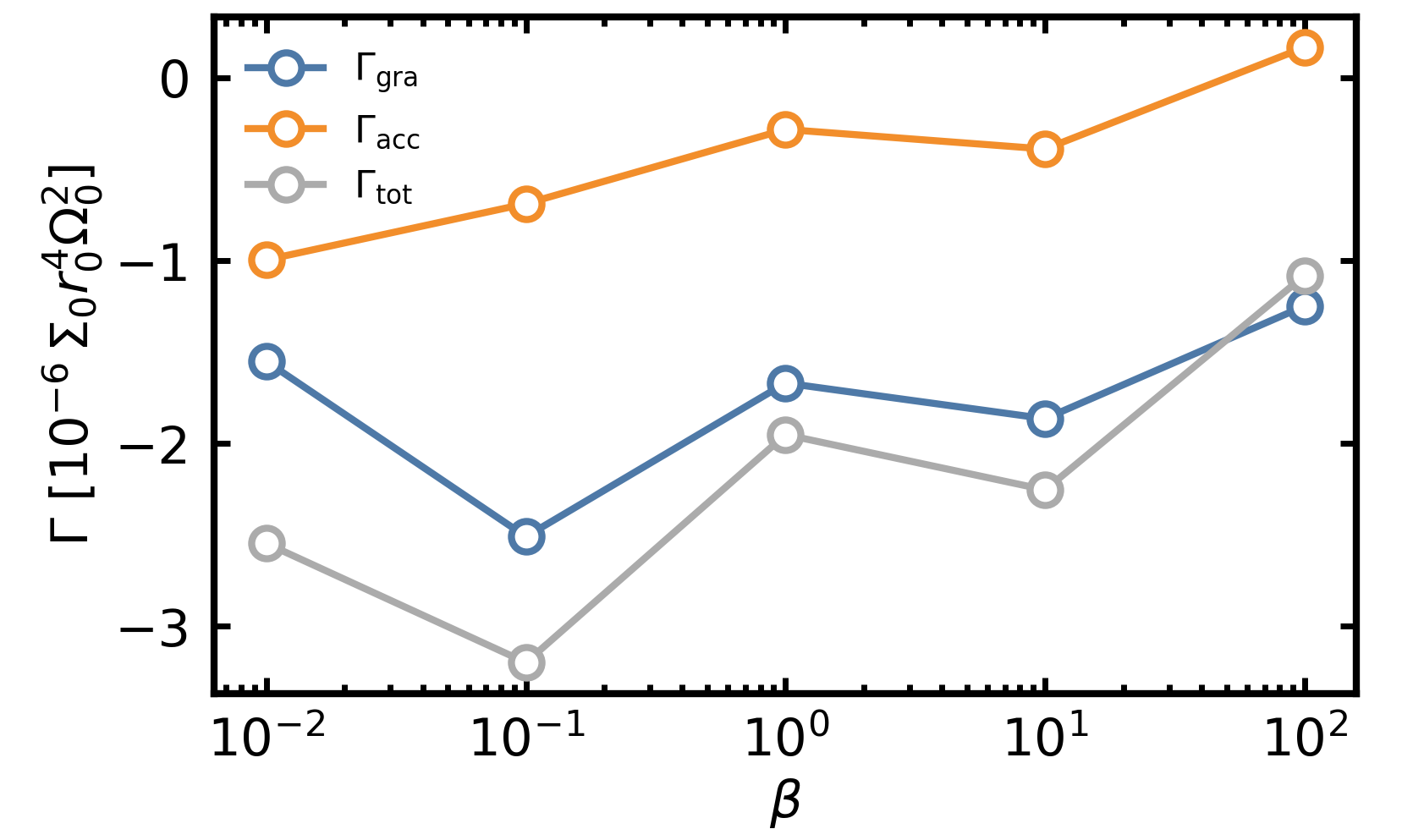}
    \caption{\added[id=R4]{Torque exerted on the planet for different cooling times, $\beta$. We set $q=10^{-3}$. Blue, orange, and gray circles show the gravitational, accretion, and total torques, respectively. The torques were computed using density and velocity fields averaged over the last 100 orbits ($900$--$1000\,t_0$).}}
    \label{fig:torque_vs_beta_q1e-03_b1e-02-1e+02}
\end{figure}

\added[id=R4]{
Here we assess the migration tendency by computing the torque exerted on the planet using quasi-steady density and velocity fields. 
Following \cite{li2024ConcurrentAccretionMigration} and \cite{wu2024EffectsThermodynamicsConcurrent}, we decompose the torque into gravitational and accretion torques:
\begin{align}
&\Gamma_{\rm gra} = \bm{r}_{\rm p}\times\int \rho\,\nabla\Phi_{\rm p}'\, \mathrm{d}V,\\
&\Gamma_{\rm acc} = \bm{r}_{\rm p} \times \int_{ r_{\rm cyl} < r_{\rm sink}} (\bm{v} - \bm{v}_p)\, \mathrm{d}\dot{M}_{\rm acc}
\end{align}
where $\Phi_{\rm p}'$ is the planetary gravitational potential including the planet-induced indirect term, $\bm{v}$ and $\bm{v}_{\rm p}$ are the gas and planetary velocities in the inertial frame, and $\mathrm{d}\dot{M}_{\rm acc}=\rho\,\mathrm{d}V/t_{\rm sink}$.
The total torque is defined as $\Gamma_{\rm tot}=\Gamma_{\rm gra}+\Gamma_{\rm acc}$.
The torque in code units is given by $\tilde{\Gamma}=\Gamma/(\Sigma_0r_0^4\Omega_0^2)$.
}

\added[id=R4]{
We find that the total torque is negative for all cooling times considered here, suggesting that a released planet would initially migrate inward (Fig.~\ref{fig:torque_vs_beta_q1e-03_b1e-02-1e+02}).
The total torque is dominated by the gravitational torque, which depends only weakly on $\beta$. 
In contrast, the accretion torque becomes less negative with increasing $\beta$ and turns positive at $\beta=10^2$, although it is not sufficient to reverse the sign of the total torque.}

\added[id=R4]{
From the measured torques, we found that the time required for the planet to migrate by $2\,R_{\rm H}$, corresponding approximately to the radial distance to the accretion bands (Fig.~\ref{fig:accretion_bands}), is much longer than the simulation time.
Here the migration rate of the planet’s semimajor axis can be expressed as \citep{li2024ConcurrentAccretionMigration,wu2024EffectsThermodynamicsConcurrent}
\begin{align}
 \frac{\dot{r}_0}{r_0}=2\,\tilde{\Gamma}_{\rm tot}\Bigg(\frac{q}{10^{-3}}\Bigg)^{-1}\Bigg(\frac{\Sigma_0r_0^2/M_\ast}{10^{-3}}\Bigg)\,\Omega_0.
\end{align}
We obtained approximately $3\times10^4$ orbits for $\beta=10^{-2}$ and $8\times10^4$ orbits for $\beta=10^2$. 
Orbital migration is therefore unlikely to significantly modify the local disk environment over the duration of these simulations.
}

\added[id=R4]{
Our results with $\alpha=10^{-3}$ are broadly consistent with \citet{li2024ConcurrentAccretionMigration}, who found that accreting planets in locally isothermal disks tend to migrate inward at viscosities below $\alpha\sim3\times10^{-3}$, while migrating outwards at higher viscosities.
The 2D $\beta$-cooling simulations of \citet{wu2024EffectsThermodynamicsConcurrent} finds a transition from outward to inward migration around $\beta\sim1$ for simulations with $\alpha=10^{-2}$. 
Quantifying the combined dependence of migration on cooling time and viscosity in 3D requires a dedicated parameter survey and is left for future work.
}

\end{appendix}
\end{document}